\documentclass[prb,aps, twocolumn, amsmath,amssymb,superscriptaddress]{revtex4}
\usepackage{float}
\usepackage{amsmath}
\usepackage{amssymb}
\usepackage{amsfonts}
\usepackage{euscript}
\usepackage{enumerate}
\usepackage{hhline}
\usepackage{pslatex}
\usepackage{tabularx}
\usepackage{color}

\usepackage{graphicx}
\usepackage{dcolumn}
\usepackage{bm}
\usepackage[sort&compress]{natbib}

\newcommand{\bra}[1]{\left< #1 \right|} 

\newcommand{\ket}[1] {| #1 \rangle}
\makeatletter
\renewcommand{\@biblabel}[1]{#1. }
\renewcommand{\@dotsep}{500}
\renewcommand{\@pnumwidth}{0em}
\renewcommand{\l@figure}[2]{

\@dottedtocline{1}{1.5em}{2em}{Figure #1}{}\vspace{15pt}}

\begin{document}

\title{Coherent coupling between radio frequency, optical, and acoustic waves in piezo-optomechanical circuits}

\author{Krishna C. Balram}\email{krishna.coimbatorebalram@nist.gov}
\affiliation{Center for Nanoscale Science and Technology, National
Institute of Standards and Technology, Gaithersburg, MD 20899,
USA}\affiliation{Maryland NanoCenter, University of Maryland,
College Park, MD 20742, USA}
\author{Marcelo I. Davan\c co}
\affiliation{Center for Nanoscale Science and Technology, National
Institute of Standards and Technology, Gaithersburg, MD 20899,
USA}
\author{Jin Dong Song}
\affiliation{Center for Opto-Electronic Materials and Devices Research, Korea Institute of Science and Technology, Seoul 136-791, South Korea }
\author{Kartik Srinivasan} \email{kartik.srinivasan@nist.gov}
\affiliation{Center for Nanoscale Science and Technology, National
Institute of Standards and Technology, Gaithersburg, MD 20899, USA}

\date{\today}

\begin{abstract}
\textbf{The interaction of optical and mechanical modes in nanoscale optomechanical systems has been widely studied for applications ranging from sensing to quantum information science.  Here, we develop a platform for cavity optomechanical circuits in which localized and interacting 1550~nm photons and 2.4~GHz phonons are combined with photonic and phononic waveguides. Working in GaAs facilitates manipulation of the localized mechanical mode either with a radio frequency field through the piezo-electric effect, or optically through the strong photoelastic effect. We use this to demonstrate a novel acoustic wave interference effect, analogous to coherent population trapping in atomic systems, in which the coherent mechanical motion induced by the electrical drive can be completely cancelled out by the optically-driven motion.  The ability to manipulate cavity optomechanical systems with equal facility through either photonic or phononic channels enables new device and system architectures for signal transduction between the optical, electrical, and mechanical domains.}
\end{abstract}

 \maketitle

\maketitle
The interaction of optical and mechanical degrees of freedom in chip-based, nanoscale systems has been studied in many contexts, ranging from cavity optomechanical systems~\cite{ref:Favero_Karrai_optomechanics_review,ref:Aspelmeyer_RMP,ref:Metcalfe_optomechanics} in which localized optical and mechanical modes are coupled, to waveguiding geometries in which propagating photons and phonons interact, as in stimulated Brillouin scattering~\cite{ref:Pant_Eggleton_SBS,ref:Rakich_Nat_Comm,ref:VanLaer_Baets_SBS} (the latter has also been studied in travelling-wave microresonators~\cite{ref:Matsko_SAW_WGM,ref:Bahl_SBS_resonator,ref:Vahala_grp_uwave_Brillouin}). Recently there has been interest in combining the exquisite motion sensitivity of cavity optomechanical systems with the radio frequency (RF) signal processing functionality of electromechanical systems to enable wavelength conversion between microwave and optical domains~\cite{ref:Lehnert_Regal_uwave_optical}, electrostatically-actuated optomechanical cavities for sensing and feedback cooling~\cite{ref:Winger_Painter_electromechanical,ref:Miao_cavity_OM_sensing}, and piezo-optomechanical cavities~\cite{ref:Bochmann_Cleland_NatPhys,ref:Fong_Tang_uwave_assisted} in which mechanical motion is driven by RF fields, with sensitive readout and coherent interference effects observable in the optical domain.

In this work, we develop a platform that combines the strong interaction achievable in a nanoscale cavity optomechanical system with coherent control of the localized mechanical mode along both the optical channel~\cite{ref:Weis_Kippenberg,ref:safavi-naeini4} and through coupling to propagating phonons using acoustic waveguides.  Working in the GaAs material system allows us to combine its photoelastic effect, which strongly couples localized optical and mechanical modes~\cite{ref:Baker2014photoelastic,ref:Balram_GaAs_MB_PE}, with its piezoelectric behavior, which enables the generation of surface acoustic waves from an RF drive~\cite{ref:Santos_SAW_photonics}.  We develop phononic crystal waveguides that guide acoustic waves that couple to the 2.4~GHz breathing mode of nanobeam optomechanical crystal cavities~\cite{ref:eichenfield2,ref:chan_optimized_OMC,ref:Balram_GaAs_MB_PE} and are sourced by RF-driven interdigitated transducers (IDT). Phase-sensitive optical measurements confirm the ability to initialize the mechanical resonator with arbitrary amplitude and phase, and to detect an average coherent intracavity phonon population of less than one. Finally, we take advantage of an optomechanical coupling rate $g_{0}/2\pi~\approx~$1.1~MHz that is one to two orders of magnitude larger than that achieved in other piezo-optomechanical systems~\cite{ref:Bochmann_Cleland_NatPhys,ref:Fong_Tang_uwave_assisted} to demonstrate an acoustic interference effect analogous to atomic coherent population trapping, in which the RF-driven coherent motion of the mechanical resonator is cancelled by the optically-induced motion, and vice versa.  Coherent manipulation of the mechanical resonator through both optical and electrical channels with equal facility, combined with on-chip routing provided by phononic crystal waveguides, enables new applications of integrated electro-optomechanical systems in signal transduction and sensing.

\noindent \textbf{Coupling Propagating and Localized Phononic Modes}
\label{sec:cavity_OM_circuits}
\normalsize

\begin{figure*}
\begin{minipage}[c]{0.6\linewidth}
\includegraphics[width=\linewidth]{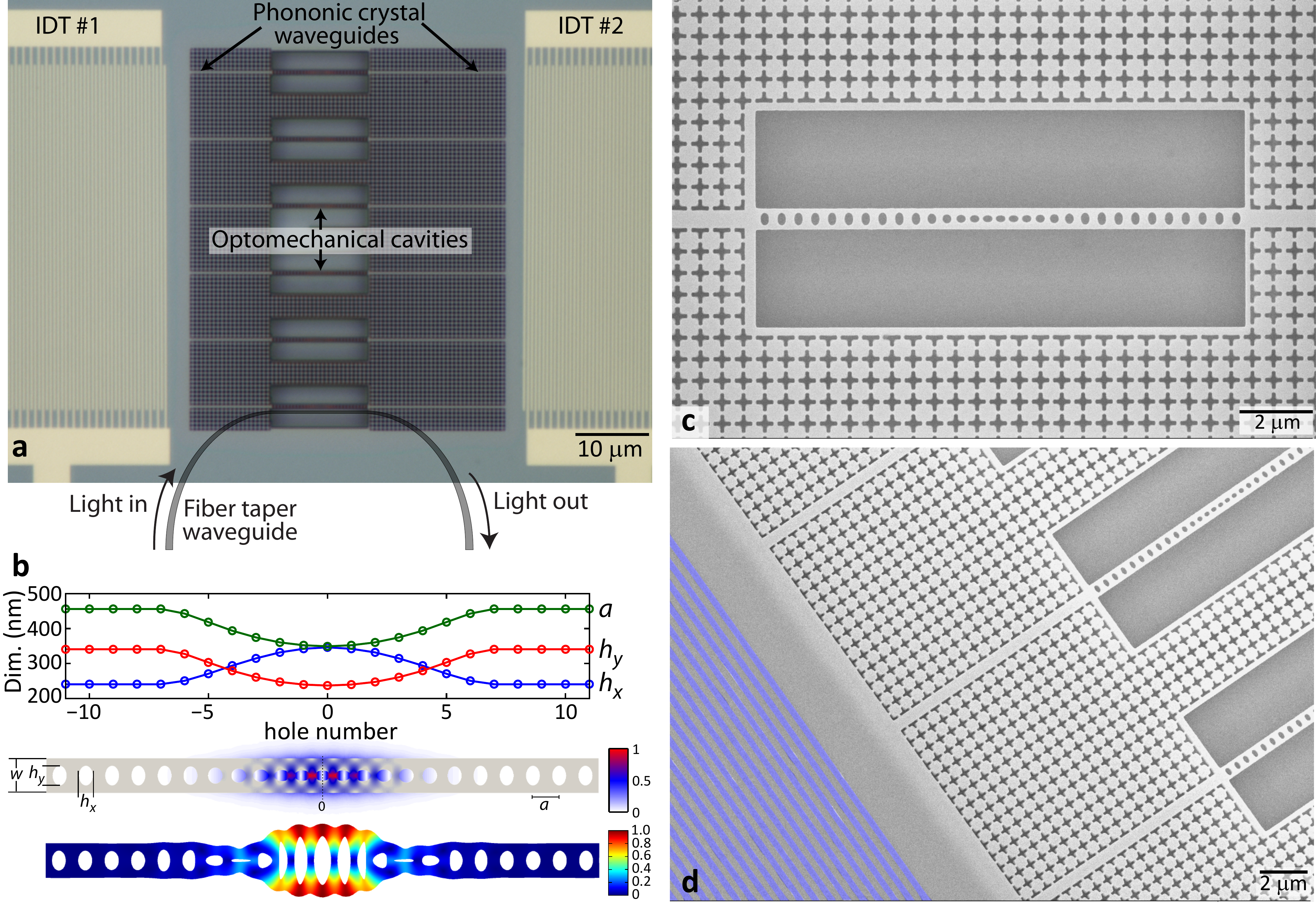}
\end{minipage}\hfill
\begin{minipage}[c]{0.38\linewidth}
\caption{\textbf{Piezo-optomechanical circuits.} (a) Optical micrograph of the device layout.  An array of optomechanical cavities, each
coupled to input and output phononic crystal waveguides, is placed between inter-digitated transducers (IDTs) that excite and/or detect surface
acoustic waves. Optical coupling is done using a fiber taper waveguide, shown schematically in gray. (b) Optomechanical crystal cavity design. (Top) Variation of the lattice constant, major axis diameter, and minor axis diameter of the elliptical holes along the nanobeam; (Middle) Simulated
normalized electric field amplitude for the 1550~nm optical mode. (Bottom) Simulated normalized displacement amplitude for the 2.4~GHz mechanical breathing mode. (c) Scanning electron microscope (SEM) image of a nanobeam optomechanical crystal cavity, including the adjacent input and output
phononic crystal waveguides. (d) SEM image showing the transition from the input IDT (shaded in blue) to the phononic crystal waveguide
and nanobeam cavity.}
\label{fig:Fig1}
\end{minipage}
\end{figure*}

\noindent Figure~\ref{fig:Fig1}(a) shows an overview of our optomechanical circuit platform~\cite{ref:piezo_optomechanical_circuits_note}. The system uses an inter-digitated transducer (IDT) to convert an applied RF drive to a propagating surface acoustic wave via the piezoelectric effect. This wave is routed using a phononic crystal waveguide and is butt-coupled to a nanobeam optomechanical crystal cavity (Fig.~\ref{fig:Fig1}(c)-(d)). The optical interface to the cavity is a fiber taper waveguide, which is used to inject and extract light from the device.  Finally, a second IDT can electrically detect surface acoustic waves out-coupled from the nanobeam cavity, or act as a second source of surface acoustic waves that couple to the cavity via a second phononic crystal waveguide.

Figure~\ref{fig:Fig1}(b) shows the optomechanical crystal cavity design, which consists of an array of elliptical air holes in a suspended nanobeam~\cite{ref:Balram_GaAs_MB_PE}.  The quadratic grade of the lattice constant and major and minor axis diameters of the elliptical holes co-localizes 1550~nm photons and 2.4~GHz phonons to a micrometer length scale. The optomechanical coupling rate $g_{0}/2\pi$, which quantifies optical cavity frequency shift due to zero-point motion, is $\approx$1.1~MHz and is predominantly due to the photoelastic effect.  This is amongst the highest reported for a cavity optomechanical system, and enables strong dynamic back-action so that the optical field can coherently manipulate the localized mechanical mode. For coherent control of the localized mechanical breathing mode in the RF domain, we need mechanisms to convert the RF voltage into a propagating acoustic wave and a waveguide geometry to route these waves and couple to the nanobeam cavity. We begin by considering the acoustic wave generation.

\begin{figure*}
\begin{center}
\includegraphics[width=0.8\linewidth]{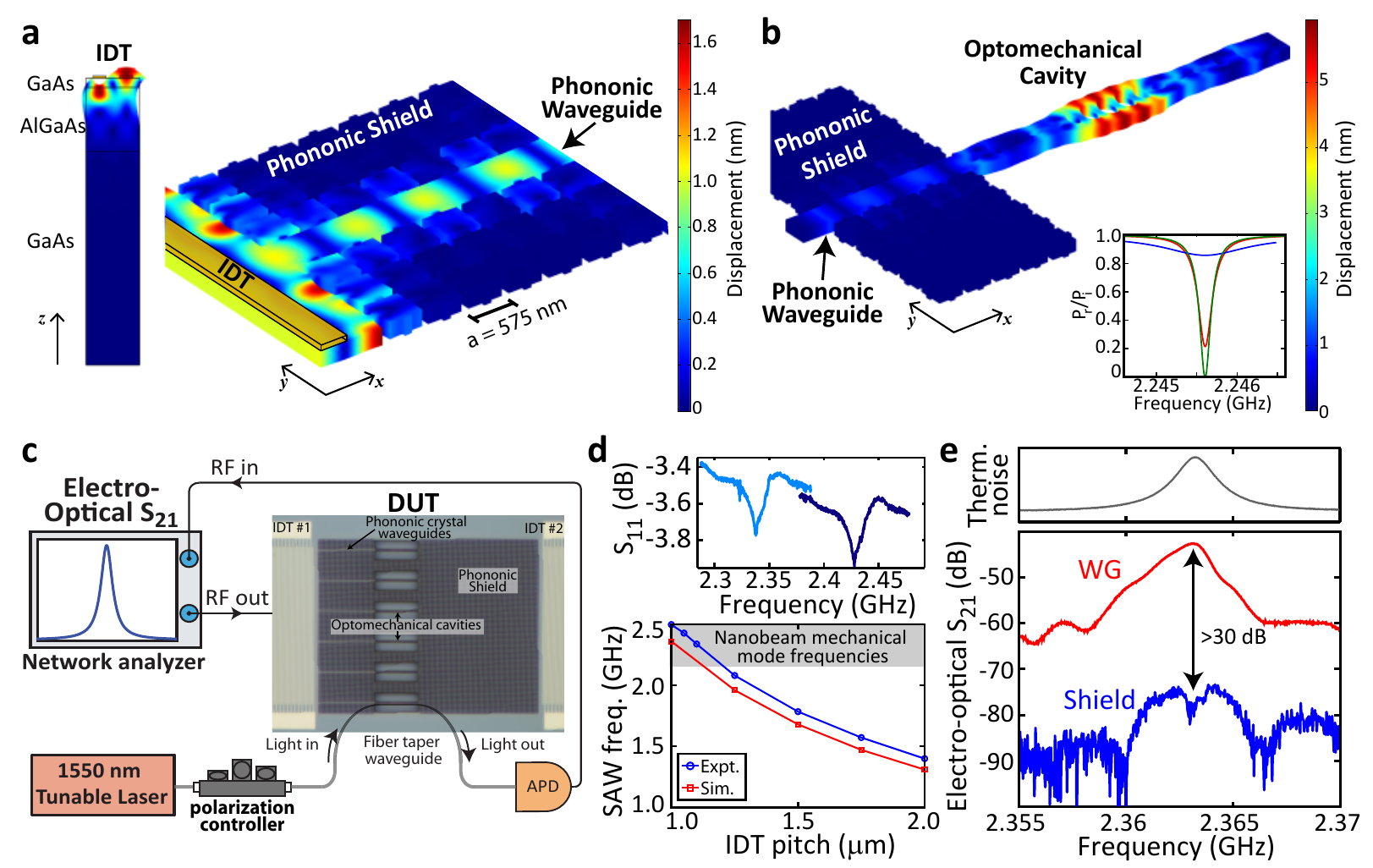}
\caption{\textbf{Coupling between propagating and localized phononic modes.} (a) Finite element method (FEM) simulation of the surface acoustic wave generated by an input IDT (left) and its excitation of the propagating mode of a phononic crystal waveguide (right). (b) FEM simulation of the coupling between the phononic crystal waveguide and breathing mode of the nanobeam optomechanical crystal cavity.  The inset shows the calculated acoustic reflection spectrum, for intrinsic mechanical quality factors of $1.5{\times}10^3$ (blue), $1.5{\times}10^4$ (red), and $2.5{\times}10^4$ (green), respectively.  (c) Experimental setup for optical readout of the nanobeam cavity coherent mechanical motion.  The cavity is either driven through a phononic crystal waveguide excited by IDT 1, or through the phononic shield excited by IDT 2. APD = avalanche photodiode.  (d) (Top) IDT reflection spectrum ($S_{11}$) for two different finger spacings. (Bottom) IDT resonance wavelength as a function of finger spacing, including both experimental results (blue circles) and FEM simulations (red squares).  (e) (Top) Thermal noise spectrum of the nanobeam cavity's localized mechanical mode.  The y-axis span is 10~dB. (Bottom) Coherently detected motion of the nanobeam mechanical cavity as a function of RF frequency, both when the system is driven through a phononic crystal waveguide (red) and through the phononic shield (blue).}
\label{fig:Fig2}
\end{center}
\end{figure*}

Surface acoustic waves (SAWs) have long been used for applications in signal processing and communication~\cite{ref:campbell_saw_review,ref:campbell_SAW_book}. Although they exist on the surface of every material, piezoelectric materials are used most extensively because of the ease of SAW excitation using IDTs, which are interleaved metal fingers biased with opposite polarity~\cite{ref:piezo_optomechanical_circuits_note}. To take into account the multilayer metal electrode and underlying epitaxial layer structure (GaAs on Al$_{0.7}$Ga$_{0.3}$As), the IDT resonance frequency is determined from finite element method (FEM) simulations. Figure~\ref{fig:Fig2}(a) shows the simulated SAW cross-section.

To route acoustic energy to the nanobeam optomechanical cavity, we use phononic crystal waveguides. The phononic crystal design is adapted from the periodic cross-structure~\cite{ref:Safavi-Naeini1}, which supports a complete bandgap for acoustic waves~\cite{ref:piezo_optomechanical_circuits_note}.  The waveguide is a line defect geometry created by removing one row of crosses, resulting in a structure that supports laterally-confined, propagating acoustic modes.  Figure~\ref{fig:Fig2}(a) shows an FEM simulation of an acoustic wave excited by an IDT and propagating through the waveguide. The phononic crystal confines the mechanical energy to the line defect region, with little displacement beyond a couple of periods transverse to the propagation direction.

We use FEM simulations to understand the coupling between the localized and propagating mechanical modes.  Figure~\ref{fig:Fig2}(b) shows the results of a simulation in which the waveguide mode at 2.25~GHz is launched into the nanobeam cavity, with excitation of the resonant mechanical breathing mode clearly observed.  This coupling can be understood quantitatively from the cavity's acoustic reflection spectrum~\cite{ref:piezo_optomechanical_circuits_note}.  The finite intrinsic $Q_{\text{m}}$ of the cavity in fabricated devices is modeled by a small imaginary component in the Young's modulus of the material, with the magnitude adjusted for different $Q_{\text{m}}$. The reflection spectra (Fig.~\ref{fig:Fig2}(b) inset) show the expected Lorentzian dip as the waveguide mode frequency sweeps over the cavity, with the depth determined by the coupling rate relative to the cavity intrinsic loss rate.  For $Q_{\text{m}}=1500$ (typical for our devices), the reflection contrast is limited to $\approx$~15~$\%$, while an improved intrinsic $Q_{\text{m}}$ of $2.5{\times}10^4$ would enable critical coupling.  Alternately, the transition between waveguide and cavity, or the waveguide geometry itself, can be tailored to suitably increase the coupling rate.

Experimentally, the IDT frequency response is measured by an electrical $S_{11}$ (reflection) measurement. The top panel of Fig.~\ref{fig:Fig2}(d) shows the $S_{11}$ spectra for IDTs with pitch of the electrode fingers of $1.05~\mu${m} and $1.1~\mu${m}. Optimal loading of the nanobeam's localized mechanical mode by the RF drive occurs when the mode frequency lies within the IDT bandwidth.  The bottom panel of Fig.~\ref{fig:Fig2}(d) shows the scaling of the IDT resonance frequency with finger pitch in simulation and experiment, along with the desired region of operation shaded in gray (corresponding to the nanobeam breathing mode frequency range).  To ensure that some fabricated devices show the required spectral overlap, each IDT addresses an array of optomechanical cavities (Fig.~\ref{fig:Fig1}(a)), and the nanobeam width is varied across the array, thereby tuning the localized mechanical mode frequency with respect to a fixed IDT frequency~\cite{ref:piezo_optomechanical_circuits_note}.

While IDTs are a convenient means to generate propagating acoustic waves for coupling to optomechanical cavities, the relatively small piezoelectric transduction coefficient of GaAs~\cite{ref:campbell_saw_review} limits the efficiency with which acoustic energy can be converted to RF energy for electrical readout.  Thus, for sensitive detection of the nanobeam optomechanical cavity's motion when driven by the propagating acoustic wave, we use the strong photoelastic coupling for optical readout.  The setup is shown in Fig.~\ref{fig:Fig2}(c), where light is injected into the optomechanical cavity using an optical fiber taper waveguide, and a vector network analyzer (VNA) drives the IDT with RF energy and coherently detects the photocurrent signal transmitted past the optomechanical cavity, yielding an electro-optic $S_{21}$ measurement.  The laser wavelength is positioned on the shoulder of the optical mode ($Q_{opt}~=~36600\pm400$, fit uncertainty 95~$\%$ confidence intervals)~\cite{ref:piezo_optomechanical_circuits_note}, so that phase fluctuations induced by motion are transduced into an intensity-modulated optical signal at the output.

The bottom panel of Fig.~\ref{fig:Fig2}(e) shows the measured electro-optic $S_{21}$ in two instances.  In the first case (red curve), a phononic crystal waveguide butt-coupled to the optomechanical cavity is adjacent to the RF-driven IDT (IDT 1, Fig.~\ref{fig:Fig2}(c)), and a pronounced peak $\approx$~20~dB above the background is observed at the nanobeam cavity's localized mechanical mode frequency, as confirmed by direct detection of its thermal noise spectrum (top panel of Fig.~\ref{fig:Fig2}(e)). In the second case (blue curve), an unperturbed phononic shield spans the region between the cavity and RF-driven IDT (IDT 2, Fig.~\ref{fig:Fig2}(c)).  Here, the electro-optic $S_{21}$ is suppressed by more than 30~dB relative to the first case, indicating that the phononic shield effectively blocks the transmission of acoustic energy from the IDT to the cavity.

\noindent \textbf{Coherent Mechanical State Preparation and Readout}
\label{sec:phonon_number_Vpi}
\normalsize

\begin{figure*}
\begin{center}
\includegraphics[width=0.8\linewidth]{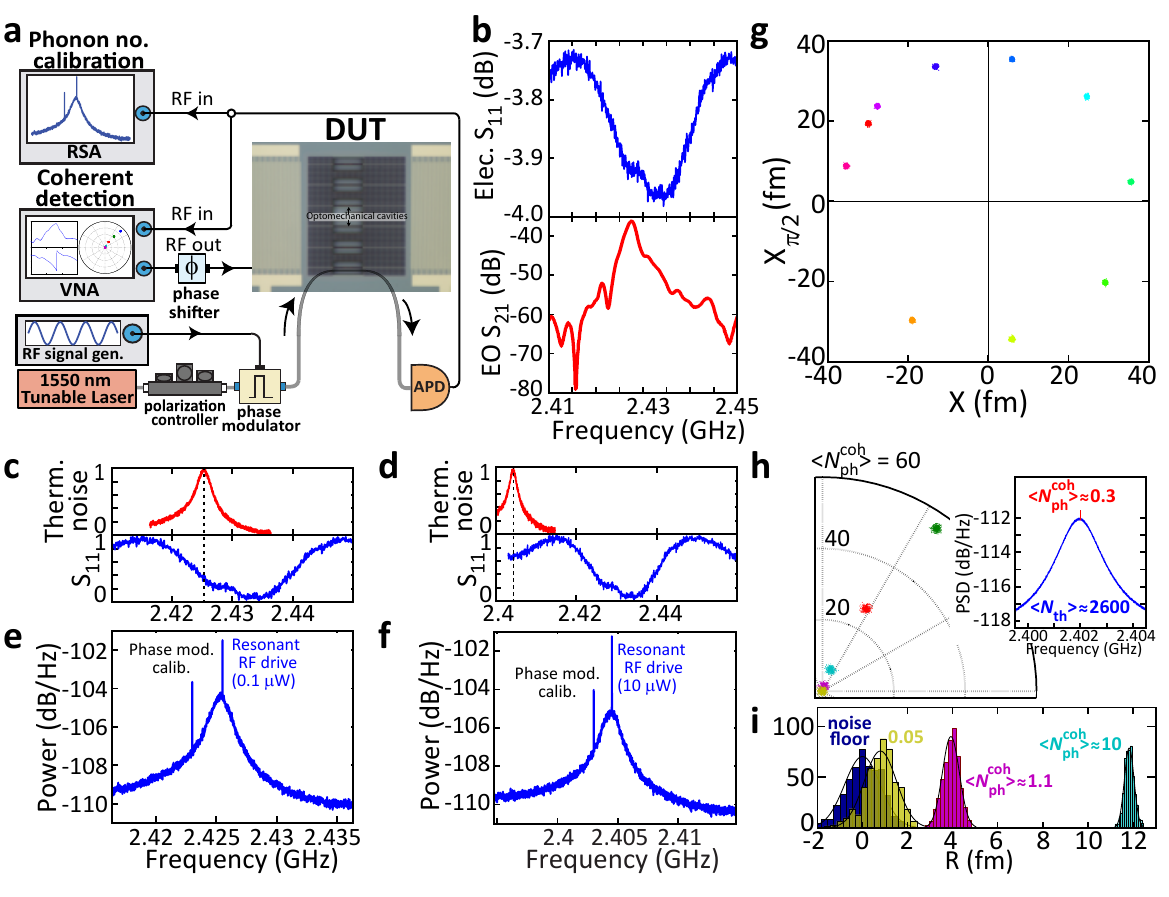}
\caption{\textbf{Acousto-optic modulation and coherent phonon detection.} (a) Setup for determining the average coherent intracavity phonon number $N^{\text{coh}}_{\text{ph}}$ in the mechanical cavity.  The coherent motion due to the RF drive is compared to the thermally-driven motion of the cavity. DUT = device under test, VNA = vector network analyzer, RSA = real-time spectrum analyzer. (b) IDT reflection spectrum (top) and optically-detected coherent mechanical motion (bottom) for an IDT and nanobeam that are spectrally aligned. (c)-(d) Thermal noise spectrum of the localized mechanical mode (red, RF turned off) and IDT electrical $S_{11}$ spectrum (blue) for an IDT and nanobeam that are spectrally (c) aligned and (d) misaligned. (e)-(f) Corresponding total photodetected spectrum with the RF drive on. $N^{\text{coh}}_{\text{ph}}=43\pm1.5$ in (e).  To achieve a similar $N^{\text{coh}}_{\text{ph}}=50\pm1.8$ in (f), the RF drive power is increased by 20~dB.  An additional tone due to phase modulation of the optical signal is used for calibrating the optomechanical coupling rate $g_{0}$. (g) Phase space diagram of the coherent mechanical motion for different values of the RF phase shifter.  Each state has $N^{\text{coh}}_{\text{ph}}=91\pm3.3$, and the VNA measurement bandwidth is 200 Hz. (h) (Left) Polar plot of $N^{\text{coh}}_{\text{ph}}$ for different RF drive amplitudes. (Right) Total photodetected spectrum with the RF drive on and attenuated to produce $N^{\text{coh}}_{\text{ph}}=0.3\pm0.01$. (i) Histogram of the the polar plot data from (h), shown as a function of motional amplitude, with $N^{\text{coh}}_{\text{ph}}$ and the measurement noise floor (obscured in the polar plot in (h)) indicated. Measurement bandwidth is 20 Hz. The uncertainty in the phonon number is due to the combined uncertainty in sample temperature, fit to the thermal noise spectrum, and RSA amplitude error and is a one standard deviation value~\cite{ref:piezo_optomechanical_circuits_note}. The power spectral density plots in (e), (f), and (h) are referenced to a power of 1 mW (0~dB = 1~mW).}
\label{fig:Fig3}
\end{center}
\end{figure*}

\noindent Having demonstrated routing, we now further quantify the phonon injection process.  Figure~\ref{fig:Fig3}(a) shows the setup used, where a VNA excites a source IDT and reads out the coherent component of the photodetected light transmitted past the optomechanical cavity, with a phase shifter now added for phase control of the injected RF tone.  In addition, a second RF signal generator drives an optical phase modulator for determining the optomechanical coupling rate $g_{0}$~\cite{ref:Gorodetsky_Kippenberg_OM,ref:Balram_GaAs_MB_PE}, and a real-time spectrum analyzer (RSA) displays the total power spectral density (PSD), consisting of the incoherent thermal motion of the mechanical resonator, the coherent motion of the resonator due to coupling from the IDT-driven phononic waveguide, and the phase modulator calibration tone.

Figure~\ref{fig:Fig3}(b) displays the electro-optic $S_{21}$ measurement of coherently-driven motion similar to Fig.~\ref{fig:Fig2}(e), but now for an optomechanical cavity connected to phononic waveguides at both its input and output (Fig.~\ref{fig:Fig3}(a)).  The results are similar, with the electro-optic $S_{21}$ showing a peak corresponding to the cavity's mechanical resonance frequency and which lies within the IDT's resonance bandwidth.  The importance of this overlap is shown in Fig.~\ref{fig:Fig3}(e)-(f), which contrasts the RF power applied to the IDT to achieve a certain coherent intracavity phonon population when the localized nanobeam mechanical mode is on-resonance (Fig.~\ref{fig:Fig3}(c)) to when it is off-resonance with the IDT (Fig.~\ref{fig:Fig3}(d)).  For this comparison, two nanobeam optomechanical cavities from the same array were chosen, so that the IDT characteristics were fixed as the nanobeam mechanical frequency changed.  Off resonance, the RF power is increased by nearly 20~dB to achieve a similar strength of the coherent tone.

We quantify the average coherent intracavity phonon number $N^{\text{coh}}_{\text{ph}}$ by comparing the strength of the coherent tone with that due to the thermally-driven motion.  That is:

\begin{eqnarray}
{N^{\text{coh}}_{\text{ph}}} = \frac{k_{\text{B}}T}{\hbar \Omega_{m}}\frac{S_{\text{coh}}(\Omega_{\text{coh}})}{S_{\text{th}}(\Omega_m)}
\end{eqnarray}

\noindent where $k_{\text{B}}$ is Boltzmann's constant, $T$ is temperature, $\Omega_{m}$ is the mechanical mode frequency, $S_{\text{coh}}(\Omega_{\text{coh}})$ is the power in the coherent tone, and $S_{\text{th}}(\Omega_m)$ is the integrated power in the thermal noise peak.  $N^{\text{coh}}_{\text{ph}}=43\pm1.5$ for Fig.~\ref{fig:Fig3}(e) and $N^{\text{coh}}_{\text{ph}}=50\pm1.8$ for Fig.~\ref{fig:Fig3}(f), quantifying the increased RF drive needed to achieve a given $N^{\text{coh}}_{\text{ph}}$ when the nanobeam mechanical resonance and IDT are spectrally misaligned.  Finally, we use the phase modulator to measure $g_{0}/2\pi~=~$1.1~MHz~$\pm~60~$kHz, corresponding well with simulations~\cite{ref:Balram_GaAs_MB_PE}, and with a one standard deviation uncertainty set by the uncertainty in the phase modulator $V_{\pi}$~\cite{ref:piezo_optomechanical_circuits_note}.

Another way to characterize the device is to consider it as an acousto-optic phase modulator (the comparison is exact in the sideband-resolved regime, where the mechanical mode frequency exceeds the optical cavity linewidth) and determine its $V_{\pi}$. For the device in Fig.~\ref{fig:Fig3}(c), $V_{\pi}\approx$ 652 mV~\cite{ref:piezo_optomechanical_circuits_note}.  This can be reduced to $V_{\pi}<~100$~mV by optimizing the path from the IDT to the optomechanical cavity and minimizing reflections and scattering at interfaces.  We note the qualitative difference between these acousto-optic modulators and previous demonstrations, in which the propagating acoustic waves were not coupled to localized mechanical resonances~\cite{ref:Santos_SAW_photonics,ref:Tadesse_Li_SAW_AOMs}. Use of a localized mechanical resonance significantly enhances the modulation efficiency without creating a significant bandwidth restriction beyond that already imposed by the IDT.

To demonstrate coherent control of the resonator, we would like to transfer both the amplitude and phase of the applied RF voltage onto the cavity displacement.  Figure~\ref{fig:Fig3}(g) plots the in-phase and quadrature components of the photodetected signal on-resonance for a variety of different phases applied to the IDT.  Here, the data is displayed as a function of average coherent mechanical displacement, where the calibration procedure described in the previous paragraph is used to determine $N^{\text{coh}}_{\text{ph}}$, and is converted to a displacement:

\begin{eqnarray}
N^{\text{coh}}_{\text{ph}} = \frac{1}{2}\left({\frac{\alpha_\text{cav}}{x_\text{zpf}}}\right)^2
\end{eqnarray}

\noindent where $\alpha_\text{cav}$ is the magnitude of the cavity displacement and $x_\text{zpf}=\sqrt{\frac{\hbar}{2m_{\text{eff}}\Omega_{m}}}$ is the zero-point motional amplitude ($m_{\text{eff}}~\approx$~0.5~pg is the simulated effective mass).  The data shows that, by varying the amplitude and phase of the applied RF signal, we can imprint a corresponding amplitude and phase on the localized mechanical mode, and thus initialize it at an arbitrary location in phase space.  These measurements confirm that coherence is preserved through the process of converting an applied RF voltage to an acoustic wave that propagates through a phononic crystal waveguide and couples to the localized mode of the optomechanical crystal cavity. By injecting propagating acoustic waves with similar amplitudes and varying phase difference through two phononic waveguides that excite the same optomechanical cavity, we can optically read out constructive and destructive interference effects occurring within the mechanical resonator~\cite{ref:piezo_optomechanical_circuits_note}.

The combination of coherent detection and the large optomechanical coupling strength enables measurement of a weak coherent intracavity phonon population on top of the $\approx$~2600 thermal phonons in the cavity.  In Fig.~\ref{fig:Fig3}(h), we reduce the amplitude of the RF drive and measure the mechanical motion on-resonance, displayed as a polar plot in terms of the phonon number and phase angle.  Figure~\ref{fig:Fig3}(i) shows the data plotted against motional amplitude for the three smallest RF drive amplitudes, where for each the data has been histogrammed into 15 equally spaced bins.  $N^{\text{coh}}_{\text{ph}}~\approx~$1.1 is well-separated from the measurement noise floor (RF drive set to zero), and $N^{\text{coh}}_{\text{ph}}$ as low as $\approx$~0.1 is resolvable. We also display the PSD for a measurement with $N^{\text{coh}}_{\text{ph}}<1$ (Fig.~\ref{fig:Fig3}(h) inset).

\noindent \textbf{Coherent Interactions between RF, Optics, and Mechanics}

\noindent The focus of our work so far has been establishing phononic crystal waveguides that are sourced by IDTs and effectively couple acoustic energy to the localized mechanical mode of a nanobeam optomechanical cavity.  In these experiments, large $g_{0}$ enables sensitive optical readout of the mechanical motion, but was not used to manipulate the mechanical resonator.  We next consider experiments in which the mechanical resonator is simultaneously excited using both RF signals (via IDTs and phononic crystal waveguides) and optical fields.  This highlights a key feature of the GaAs optomechanical circuit platform, which is that the motion of the mechanical resonator can be manipulated with similar facility along either the optical or RF channel.

\begin{figure*}
\begin{center}
\includegraphics[width=0.8\linewidth]{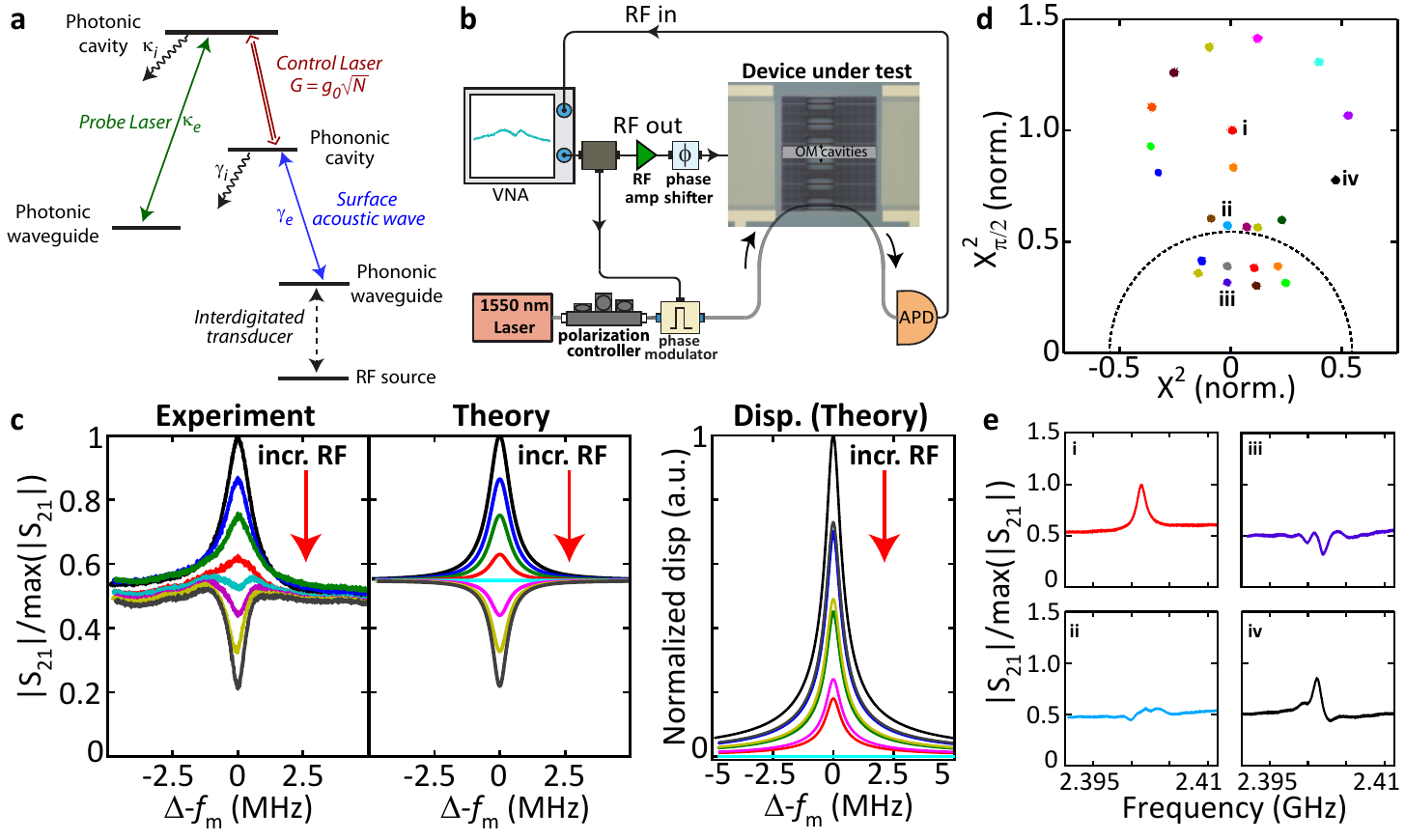}
\caption{\textbf{Acoustic wave interference analogous to coherent population trapping.} (a) Schematic level diagram indicating how the localized phononic cavity is populated either through the optical channel (via photoelastic coupling to the photonic cavity) or through RF-generated propagating acoustic waves. These two pathways can coherently interfere and cancel the mechanical motion, in an effect analogous to atomic coherent population trapping.  (b) Experimental setup. One part of the VNA output is sent to an optical phase modulator, after which interference between the generated optical sideband and carrier drives the phononic cavity optically.  The other part of the VNA output is amplified and sent through an RF phase shifter into the input IDT to generate a propagating acoustic wave to couple to the phononic cavity.  Readout of the cavity's coherent mechanical motion is done optically. (c) Experimental data (left) and theoretical calculations (center) of the coherent mechanical motion as a function of RF drive frequency and for increasing RF power.  Pure optomechanically-driven motion (RF~=~0) is shown in the black curves, acoustic coherent population trapping is shown in cyan, and RF-dominated motion is shown in dark gray.  The rightmost set of curves shows the corresponding theoretical calculations for the cavity displacement amplitude.  (d) Phase space diagram for different values of the RF amplitude and phase.  The black dashed semicircle is the far off-resonance background. The four numbered points, with frequency response curves shown in (e), correspond to (i) pure optomechanically-driven motion, (ii) acoustic coherent population trapping, (iii) RF-dominated motion, and (iv) Fano lineshape behavior.  The measurement bandwidth in (d)-(e) is 200~Hz.}
\label{fig:Fig4}
\end{center}
\end{figure*}

Figure~\ref{fig:Fig4}(a) shows a schematic level diagram indicating the couplings and decay channels of the different components in our system.  If we ignore the states on the right hand side of the diagram that depict the phononic waveguide, its feeding from an IDT, and coupling to the phononic cavity, we recover the characteristic $\Lambda$-system configuration in which optomechanically induced transparency (OMIT)~\cite{ref:Agarwal_Huang_OMIT,ref:Weis_Kippenberg,ref:safavi-naeini4} has been observed in a wide variety of systems. Adding the new channels provides us with another knob to coherently control and probe the system properties and observe new phenomena. The interference effect we describe in this section occurs in the phononic cavity, which can be driven optically via the beating of the probe and control lasers that are detuned from each other by the localized mechanical mode frequency (as in typical OMIT), or acoustically through the phononic waveguide. These two pathways can be tuned to have equal amplitudes and opposite phases, resulting in a cancellation of the coherent motion of the mechanical resonator.  This effect is analogous to coherent population trapping in atomic systems~\cite{ref:Arimondo_CPT,ref:Khan_CPT_EIT}.

Starting with the coupled equations of cavity optomechanics, one can derive an expression for the cavity displacement amplitude $\beta_{d+}$ (see supplementary material~\cite{ref:piezo_optomechanical_circuits_note}):

\begin{eqnarray}
\beta_{d+}  = \frac{-ig_{0}(\alpha_{0}\alpha^{*}_{d-})-\sqrt{\frac{\gamma_{e}}{2}}\beta_{in,0}e^{i\psi}}{i(\Omega_{m}-\Omega_{d})+\frac{\gamma_{i}}{2}}
\end{eqnarray}

The first term in the numerator corresponds to the optical drive term due to beating of the intracavity control ($\alpha_{0}$) and probe ($\alpha_{d-}$) beams and the second term corresponds to the acoustic drive term ($\beta_{in,0}e^{i\psi}$) due to acoustic waves coupling to the cavity (intrinsic decay rate $\gamma_{i}$) through the phononic waveguide (coupling rate $\gamma_{e}$). Appropriate choice of the amplitude ($\beta_{in,0}$) and phase ($\psi$) of the acoustic drive term produces the condition for cancellation of the mechanical mode ($\beta_{d+}=0$):
\begin{eqnarray}
\sqrt{\frac{\gamma_{e}}{2}}\beta_{in,0}e^{i\psi}= \frac{-i\sqrt{\frac{\kappa_{e}}{2}}g_{0}\alpha_{0}\alpha^{*}_{in,-}}{-i(\Delta+\Omega_{d})+\frac{\kappa_{i}}{2}}
\end{eqnarray}
\noindent wherein $\alpha_{in,-}$ represents the input probe signal incident on the cavity, $\Delta$ the laser-cavity detuning, $\Omega_{d}$ the probe beam detuning from the control beam frequency, $\kappa_{e}$ the optical cavity coupling rate, and $\kappa_{i}$ the intrinsic optical cavity decay rate.

When this condition is satisfied, the coherent cavity displacement tends to zero and the cavity transmission is flat over the mechanical resonator's bandwidth (the optomechanical cavity acts as a pure optical cavity to the probe signal). The mechanical cavity effectively enters a ``dark" state for phonons analogous to the dark state for photons in coherent population trapping.  It is important to note that only the coherently driven component of the cavity displacement becomes zero, as the cavity is still driven by incoherent thermal motion and has a finite power spectral density.  We also emphasize the distinction between this effect and other interference phenomena observed in cavity optomechanical systems.  OMIT is an optical interference effect in which probe photons interfere with photons scattered by the driven mechanical resonator. A closely-related effect in piezo-optomechanical systems has been termed electromechanically-induced optical transparency~\cite{ref:Bochmann_Cleland_NatPhys} or microwave-assisted transparency~\cite{ref:Fong_Tang_uwave_assisted}, where interference still occurs in the optical domain, but the mechanical resonator is driven by an RF voltage. In these situations, which we also observe in our system, the mechanical resonator has a finite coherent displacement amplitude at all times. In contrast, during acoustic coherent population trapping, the mechanical resonator's coherent displacement amplitude is zero. Coherent wavelength conversion between two optical modes coupled to the same mechanical mode~\cite{ref:Hill_Painter_WLC,ref:Dong_Wang_OM_dark_mode,ref:Liu_yuxiang_wlc} is similar in that it can also be viewed as being mediated by an optomechanical dark mode~\cite{ref:Wang_Clerk_PRL,ref:Dong_Wang_OM_dark_mode}, although coherent mechanical motion is suppressed purely though optomechanical coupling.

Figure~\ref{fig:Fig4}(b) shows the experimental setup used to probe the coherent interaction between RF, optics, and mechanics in our system. The RF output from port 1 of a VNA is split in two and drives both an electro-optic phase modulator and the IDT.  An RF amplifier and phase shifter control the amplitude and phase of the RF signal driving the IDT with respect to that driving the phase modulator, and photodetected light transmitted past the optomechanical cavity is detected by port 2 of the VNA.  The control laser power is fixed and is blue-detuned by a mechanical frequency from the optical cavity resonance.

Figure~\ref{fig:Fig4}(c) shows a series of electro-optical $S_{21}$ measurements performed as the VNA output frequency is swept across the mechanical resonance frequency and the RF power into the IDT is increased, along with calculations of the expected $S_{21}$ spectrum and corresponding mechanical resonator displacement amplitude~\cite{ref:piezo_optomechanical_circuits_note}.  The topmost black curves are for the IDT channel turned off, and correspond to the pure OMIT case. We then increase the RF power while keeping the phase fixed to satisfy the condition for acoustic wave interference. As the RF power is increased, the OMIT peak is reduced and at a particular RF power (cyan curves in Fig.~\ref{fig:Fig4}(d)), the transmission spectrum reaches zero on-resonance and is approximately (perfectly) spectrally flat in experiment (theory).  This corresponds to complete suppression of the coherent cavity displacement and removal of the transparency window that was initially induced by the optomechanical interaction.  The inability to achieve perfect spectral flatness in experiment is attributed to the actuation of other mechanical modes that occurs electrically but not optically~\cite{ref:piezo_optomechanical_circuits_note}. If we further increase the RF power, the cavity is primarily driven by the phononic waveguide channel and the cavity displacement amplitude increases correspondingly (magenta, yellow, dark gray curves). Since the phase is set for destructive interference, the probe transmission drops and for these higher RF powers, the interference feature within the optical cavity spectrum re-appears (the system is now in the regime of electromechanically-induced optical transparency~\cite{ref:Bochmann_Cleland_NatPhys,ref:Fong_Tang_uwave_assisted}). Here, the mechanical breathing mode is in the same state as in the pure OMIT case, except that the displacement has the opposite phase and so transparency is converted to absorption.

A more complete picture is observed by measuring the in-phase and quadrature components of the on-resonance photodetected signal (Fig.~\ref{fig:Fig4}(d)). Starting with the pure OMIT condition (IDT turned off, Fig.~\ref{fig:Fig4}(e), curve i), as we increase the RF drive to the IDT, for the right choice of phase the cavity displacement approaches zero (Fig.~\ref{fig:Fig4}(e), curve ii), and the signal on-resonance is equal to its off-resonance value (Fig.~\ref{fig:Fig4}(e), dashed circle).  Further increase in the RF power moves the system into the region of induced absorption (Fig.~\ref{fig:Fig4}(e), curve iii), while other parts of phase space are accessed by varying the phase applied to the IDT. For an arbitrary phase, the response of the system is given by an asymmetric Fano lineshape (for example, Fig.~\ref{fig:Fig4}(e), curve iv).

\noindent \textbf{Conclusions}
\label{sec:discussion}
\normalsize

We have presented an integrated piezo-optomechanical circuits platform in which fiber-coupled GaAs optomechanical crystal cavities are interfaced with phononic waveguides and interdigitated transducers.  This system combines localized and interacting 1550~nm photons and 2.4~GHz phonons with excitation and readout along either the optical or mechanical channel, with the latter connected to the RF domain via IDTs.  A distinguishing feature of GaAs is that it is both piezoelectric and has a strong photoelastic effect, so that the mechanical resonator can be effectively manipulated with equal facility by either RF or optical fields.
Going forward, these capabilities may enable the storage and retrieval of either optical or acoustic pulses, while phononic routing between cavity optomechanical nodes has been envisioned for both classical and quantum device applications~\cite{ref:safavi-naeini3,ref:habraken_rabl_phonon_routers}.  Further technical development may include optimized IDTs~\cite{ref:campbell_saw_review,ref:deLima_Santos_focusing_SAWs} for increased acoustic wave generation efficiency, or use of stimulated Brillouin scattering as an optical mechanism for acoustic wave generation~\cite{ref:Pant_Eggleton_SBS,ref:Rakich_Nat_Comm,ref:VanLaer_Baets_SBS}. Developments in chip-based phononics~\cite{ref:khelif_laude_2004guiding,ref:maldovan_sound_review,ref:olsson_el_kady_phononics,ref:maldovan_sound_review,ref:hatanaka_mahboob,ref:Mohammai_Adibi_phononics} can inform future phononic circuitry to interface with the optomechanical cavities, while the incorporation of InAs/GaAs quantum dots~\cite{ref:Metcalfe_Lawall_QD_SAW,ref:Krenner_grp_SAW_PC,ref:Yeo_strain_QD}, which have been used in a number of quantum optics applications~\cite{ref:Lodahl_Stobbe_RMP} and are fully compatible with the device fabrication described here, would add qualitatively new functionality.


\newpage
\onecolumngrid \bigskip

\begin{center} {{\bf \large SUPPLEMENTARY
INFORMATION}}\end{center}

\setcounter{figure}{0}
\makeatletter
\renewcommand{\thefigure}{S\@arabic\c@figure}

\setcounter{equation}{0}
\makeatletter
\renewcommand{\theequation}{S\@arabic\c@equation}

\section{Fabrication Procedure}

The epitaxial material used in this work consists of a 220~nm thick GaAs layer on a 1.5~$\mu$m thick Al$_{0.7}$Ga$_{0.3}$As sacrificial layer. The samples were spin-coated with positive tone electron beam resist and baked at $180\,^{\circ}{\rm C}$ for 120~s. The IDT patterns were exposed in a 100~keV direct write electron beam lithography system with a beam current of 2 nA and nominal dose of 580~$\mu$C/cm$^2$. After exposure, the electron beam resist was developed using MIBK:IPA 1:3 solution (90~s). To remove residual undeveloped resist, the samples were exposed to an O$_{2}$ plasma (6.7 Pa = 50 mTorr, 75 W) for 5~s. The IDT metallization was carried out in an electron beam evaporator with the metal stack Cr (5 nm)/Pt (15 nm)/Au (30 nm) deposited in succession. The lift-off was carried out by soaking the samples overnight in acetone and using gentle sonication.

For the nanobeam overlay, the samples were spin-coated with positive tone electron beam resist and baked at $180\,^{\circ}{\rm C}$ for 120~s. The nanobeam patterns were exposed in a 100~keV direct write electron beam lithography system with a beam current of 200 pA and nominal dose of 250~$\mu$C/cm$^2$. The nanobeam patterns were aligned to the IDT using alignment marks that were exposed along with the IDTs. After exposure, the electron beam resist was developed using hexyl acetate (65~s). To remove residual undeveloped resist, the samples were exposed to an O$_{2}$ plasma (6.7 Pa = 50 mTorr, 75 W) for 10~s. The nanobeam patterns were then transferred to the underlying GaAs layer using an inductively coupled plasma reactive ion etcher with an Ar/Cl$_2$ chemistry. The electron beam resist was stripped using trichloroethylene, and the nanobeams were undercut with a timed wet etch using either concentrated (49~$\%$)  HF solution or (NH$_4$)$_2$S and dilute HF (50:1 volume dilution in H$_2$O).

\section{SAW resonance simulation}

To calculate the resonance frequency for surface acoustic waves for our material stack, we model a one wavelength width cross-section of the stack with periodic boundary conditions and solve for the resonance frequency of the stack. We used the following parameters to model Al$_{x}$Ga$_{1-x}$As with $x~=~0.7$: $e_{14}~=~-0.16-0.065x$~C/m$^{2}$, $C_{11}~=~(118.8+1.4x)$~GPa, $C_{12}~=~(53.8+3.2x)$~GPa and $C_{44}~=~(59.4-0.5x)$~GPa~\cite{ref:rumyantsev1999handbook2_SI}.  The calculated frequencies are uniformly lower than the measured frequencies, which we attribute to imperfect knowledge of the density and thickness of the metallic stack (Cr/Pt/Au) comprising the IDT.

\section{Frequency response analysis for acoustic mode coupling}

For the simulation results shown in Figure 2(b) in the main text, the surface acoustic wave was launched by applying a prescribed displacement of 1 nm (the simulation was linear in displacement, the starting displacement amplitude is arbitrary) in the $x$-direction to the end facet of the line-defect waveguide ($x$ is the direction of wave propagation). The $x$-displacement is chosen so as to launch a symmetric Lamb wave in the defect waveguide and nanobeam that has the right $z$-symmetry to couple to the nanobeam breathing mode.

To calculate the coupling efficiency shown in Figure 2(b) we define the acoustic Poynting vector ($P_{j}$)
\begin{eqnarray}
P_{j}=\frac{1}{2}Re(-T_{ij}v^{*}_{i})
\end{eqnarray}

{\noindent}where $T_{ij}$ is the stress and $v_{i}$ is the particle velocity. The reflection spectrum is calculated by doing two simulations, one with the nanobeam cavity and a second with a bare waveguide whose width is equal to that of the nanobeam cavity. The Poynting vector is calculated at the input port of the cavity and the reflected power is defined as:

\begin{eqnarray}
R=1-\frac{P_\text{trans,cav}}{P_\text{trans,wvg}}
\end{eqnarray}

{\noindent}where the transmitted powers ($P_\text{trans,cav}$ and $P_\text{trans,wvg}$) are all calculated in the direction of the propagating mode ($x$).

\section{Nanobeam breathing mode frequency dependence on width}

\begin{figure}[h]
\centerline{\includegraphics[width=0.75\linewidth]{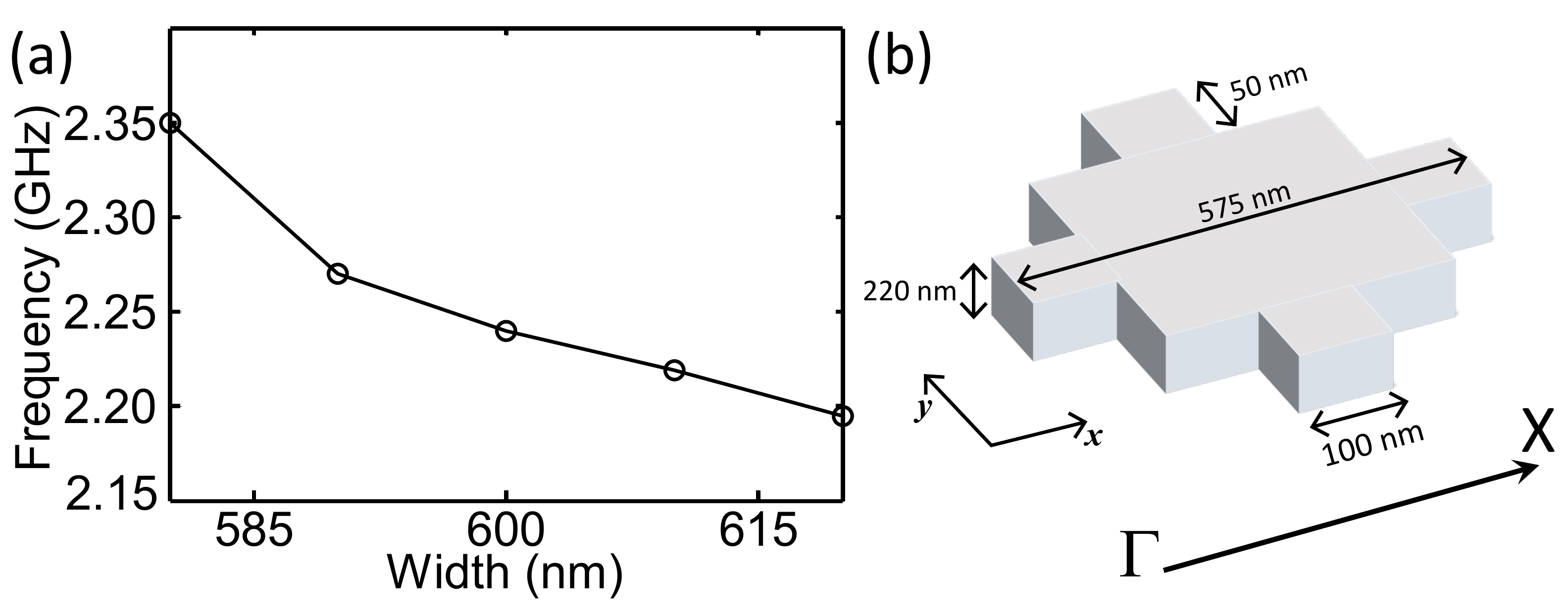}}
 \caption{(a) Calculated nanobeam optomechanical crystal cavity mechanical breathing mode frequency as a function of beam width. (b) Schematic of the unit cell of the phononic shield showing the different dimensions.}
\label{fig:FigureSI_nb_array_width_dependence}
\end{figure}

To ensure that the resonance frequency of the nanobeam optomechanical cavity's mechanical breathing mode  lies within the excitation bandwidth of the IDT, we excite an array of cavities with varying beam width with the same IDT. Figure~\ref{fig:FigureSI_nb_array_width_dependence}(a) shows the dependence of the breathing mode frequency on the width of the nanobeam and indicates that a 40 nm tuning in the width results in an approximately 155~MHz tuning in breathing mode frequency. In fabricated devices, the width of the individual beams within the nanobeam array is varied in 5 nm steps.

\section{Phononic Shield band structure}

\begin{figure}[h!]
\centerline{\includegraphics[width=0.75\linewidth]{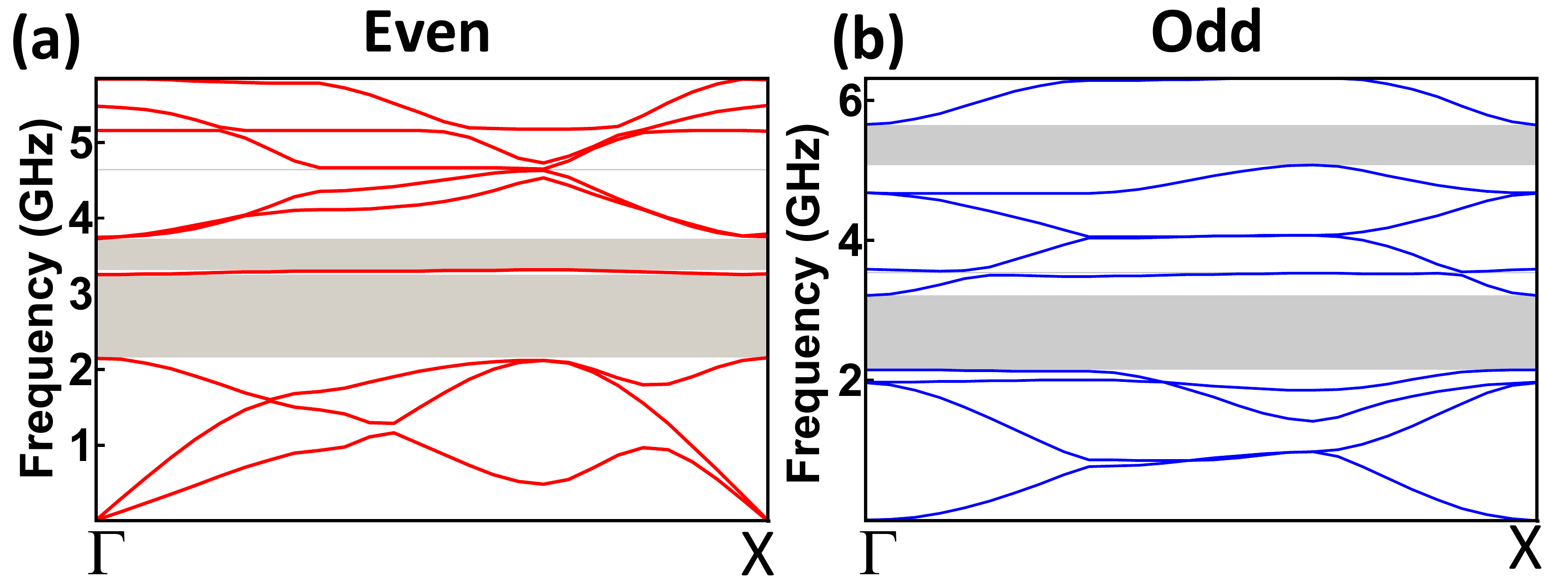}}
 \caption{Calculated phononic bandstructure for acoustic waves with (a) even and (b) odd symmetry in the out-of-plane direction ($z$), propagating along the $\Gamma$-X direction in the phononic shield. The presence of a complete bandgap for the frequencies of interest (2.25 GHz to 2.75 GHz) is clearly seen.}
\label{fig:FigureSI_Phononi_shield_bs}
\end{figure}

The dispersion diagram for acoustic wave propagation in the $\Gamma$-X direction for the phononic shield (whose unit cell is shown in Figure~\ref{fig:FigureSI_nb_array_width_dependence}(b)) is shown in Figure~\ref{fig:FigureSI_Phononi_shield_bs}. The modes of a unit cell with periodic boundary conditions in the propagation direction ($\Gamma$-X) were calculated using a finite-element solver to determine the dispersion diagram. The modes are classified as odd / even according to the symmetry in the out of plane ($z$) direction. GaAs is modelled as an orthotropic elastic material with parameters $E_{x}$ = 121.2 GPa, $E_{y}$ = 121.2 GPa, $E_{z}$ = 85.9 GPa, $\nu_{xy}$ = 0.0209, $\nu_{yz}$ = 0.4434, $\nu_{xz}$ = 0.312, $G_{xy}$ = 32.5 GPa, $G_{yz}$ = 59.4 GPa, and $G_{xz}$ = 59.4 GPa.

We note that a wide variety of phononic shield structures based on circular / square holes  have been studied~\cite{ref:Laude_Adibi_PhC_SI,ref:mohammadi2010simultaneous_SI,ref:pennec2010simultaneous_SI,ref:el-jallal_phoxonic_SI}. For the GaAs material system, achieving a complete bandgap within such geometries necessitates high hole filling fractions, and in the 2.4 GHz frequency range this requires patterning dimensions at the sub-25 nm length scale, making fabrication challenging. The advantage of the cross structure is that it allows one to achieve a complete acoustic band-gap with minimum feature size of 100 nm, which can be easily obtained by standard electron-beam lithography.

\section{Basic optomechanical spectroscopy}

\begin{figure}[h]
\centerline{\includegraphics[width=0.75\linewidth]{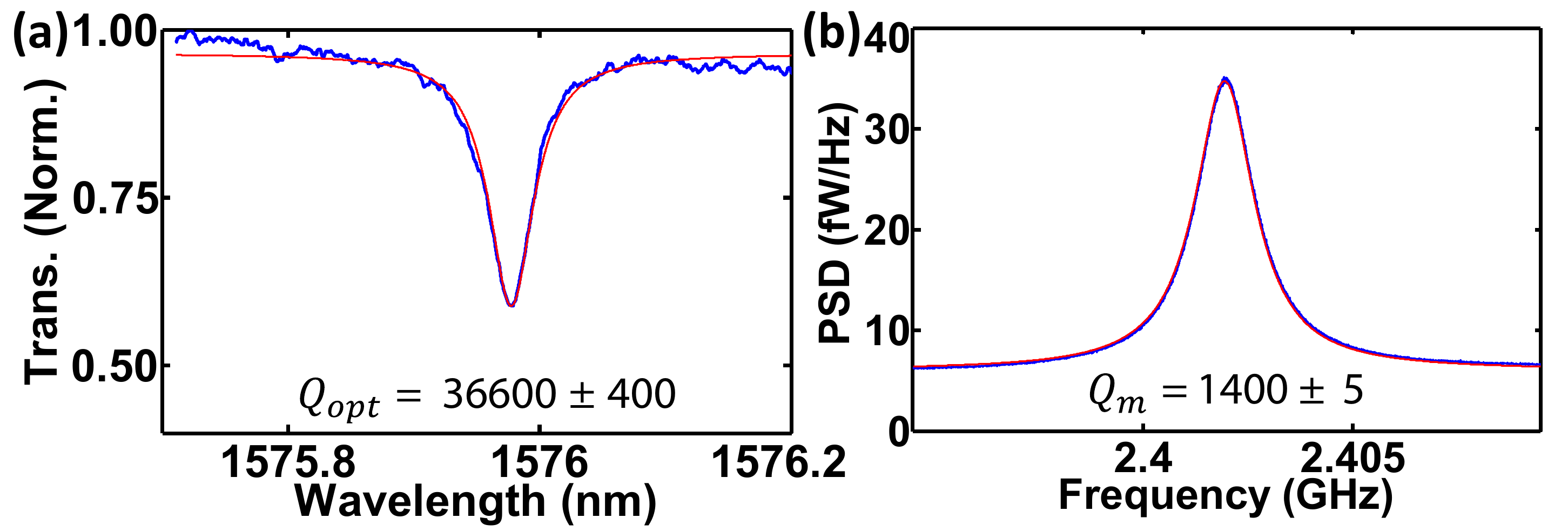}}
 \caption{(a) Transmission spectrum of the cavity optical mode. (b) Thermal noise power spectral density (PSD) of the nanobeam breathing mode (as measured on the RSA). Blue curves represent experimental data, Red: Lorentzian fits. The uncertainty in the optical and mechanical quality factors represent Lorentzian fit uncertainty, 95~$\%$ confidence intervals.}
\label{fig:FigureSI_opt_mech_mode}
\end{figure}

Figure~\ref{fig:FigureSI_opt_mech_mode}(a) shows the optical transmission spectrum of the device in Figure 3(g)-(h) and Figure 4 (main text) whose $Q_{opt}=~36600~\pm~400$ (uncertainty from the 95~$\%$ confidence interval of a Lorentzian fit to the data). Figure~\ref{fig:FigureSI_opt_mech_mode}(b) shows the corresponding thermal noise spectrum for the localized mechanical mode with $Q_{m}=~1400~\pm~5$ (uncertainty from the 95~$\%$ confidence interval of a Lorentzian fit to the data). The optomechanical coupling rate $g_{0}/{2\pi}$ for the device was measured to be 1.1~MHz$~\pm~60$~kHz using the procedure discussed below.

\section{$g_{0}$ calibration}

We use a phase modulator to calibrate the optomechanical coupling rate $g_{0}$~\cite{ref:Gorodetsky_Kippenberg_OM_SI,ref:Balram_GaAs_MB_PE_SI}. The basic idea is to relate the modulation produced by sending light through the cavity optomechanical system (which
is driven by its contact with the thermal environment) to a direct phase modulation applied with an
electro-optic phase modulator.  Because both undergo the same transduction function, the ratio of the integrated power in their photocurrent
RF spectra will be related to the ratio of $g_{0}$ and the phase modulator's modulation index.  We go through
this derivation below.

The root-mean-square (rms) thermal displacement amplitude ($\alpha_\text{thermal}$) is related to the temperature ($T$) using the equipartition theorem:%

\begin{eqnarray}
\frac{1}{2}m_\text{eff}\Omega_{m}^{2}\alpha_\text{thermal}^{2}=\frac{1}{2}k_\text{B}T
\end{eqnarray}

\noindent where $m_\text{eff}$ is the motional mass of the resonator and $\Omega_{\text{m}}$ is the
mechanical mode frequency.  We define a thermal modulation index:%

\begin{eqnarray}
\beta_\text{thermal}=\frac{\alpha_\text{thermal}g_\text{om}}{\Omega_{\text{m}}}%
\end{eqnarray}

\noindent where the optomechanical coupling parameter ($g_\text{om}$) is defined by:%

\begin{eqnarray}
\omega(\alpha)=\omega_{0}+g_\text{om}\alpha
\end{eqnarray}

\noindent with $\omega_{0}$ being the unperturbed cavity frequency.  This leads to:%

\begin{eqnarray}
\beta_\text{thermal}^{2}=\frac{k_\text{B}T}{m_\text{eff}\Omega_{\text{m}}^{2}}\frac{g_\text{om}^{2}%
}{\Omega_{\text{m}}^{2}}%
\end{eqnarray}

The optomechanical coupling rate $g_{0}$ is:

\begin{eqnarray}
g_{0}=g_\text{om}x_\text{zpf}%
\end{eqnarray}

\noindent where the (rms) amplitude of the zero point fluctuation is:%

\begin{eqnarray}
x_\text{zpf}=\sqrt{\frac{\hbar}{2m_\text{eff}\Omega_{\text{m}}}}%
\end{eqnarray}

\noindent which can be directly calculated as ${\sqrt{\bra{0} x^2 \ket{0}}}$ from the ground state wavefunction $\ket{0}$ of
the simple harmonic resonator.

We then relate $g_{0}$ to $\beta_\text{thermal}$:%

\begin{eqnarray}
\beta_\text{thermal}^{2}=\frac{2k_\text{B}T}{\hbar\Omega_{\text{m}}^{3}}g_{0}^{2}%
\end{eqnarray}

\noindent The modulation index ($\beta_\text{pm}$) of the phase modulator is defined as:%

\begin{eqnarray}
\beta_\text{pm}=\frac{\pi V_\text{sig}}{V_{\pi}}%
\end{eqnarray}

\noindent where $V_\text{sig}$ is the signal amplitude and $V_{\pi}$ is the modulator half
wave voltage.

By comparing the (integrated) powers in the cavity mechanical mode signal ($S_\text{cav}(\Omega_{\text{m}})$) and
the phase modulator signal ($S_\text{pm}(\Omega_\text{pm})$) obtained from the electronic spectrum analyzer, we get:%

\begin{eqnarray}
g_{0}^{2}=\frac{\hbar\Omega_{\text{m}}}{2k_\text{B}T}\Omega_{\text{m}}^{2}\beta_\text{pm}^{2}%
\frac{S_\text{cav}(\Omega_{\text{m}})}{S_\text{pm}(\Omega_\text{pm})}%
\end{eqnarray}

\section{Extracting cavity displacement and phonon number due to excitation via the phononic waveguide}

The modulation index due to the propagating wave $\beta_\text{saw}$ is given by:

\begin{eqnarray}
\beta_\text{saw}=\frac{\alpha_\text{cav}g_\text{om}}{\Omega_\text{saw}}
\end{eqnarray}
wherein $\alpha_\text{cav}$ is the cavity displacement (rms) due to the coupling from the propagating acoustic wave.

To extract the average coherent intracavity phonon number $N^\text{coh}_\text{ph}$, we compare the area under the coherent acoustic wave peak with the integrated area of the mechanical mode thermal noise spectrum, noting that the average thermal phonon occupation number is given by $N_{th}\approx~k_\text{B}T/{\hbar}\Omega_{\text{m}}$

\begin{eqnarray}
{N^{\text{coh}}_{\text{ph}}} = \frac{k_{\text{B}}T}{\hbar \Omega_{\text{m}}}\frac{S_{\text{coh}}(\Omega_{\text{coh}})}{S_{\text{th}}(\Omega_m)}
\end{eqnarray}

We can now relate the cavity's coherent rms displacement amplitude $\alpha_\text{cav}$ to $N_{ph}^{coh}$ using:

\begin{eqnarray}
N_\text{ph}^\text{coh}\hbar\Omega_{\text{m}} = m_\text{eff}\Omega^{2}_{m}\alpha^{2}_\text{cav}
\end{eqnarray}

{\noindent}which can be re-written as:
\begin{eqnarray}
N_\text{ph}^\text{coh} = \frac{1}{2}\left({\frac{\alpha_\text{cav}}{x_\text{zpf}}}\right)^2
\end{eqnarray}

\noindent where we have substituted $x_\text{zpf}=\sqrt{\frac{\hbar}{2m_\text{eff}\Omega_{\text{m}}}}$.

\section{Acousto-optic modulator $V_{\pi}$ calculation}

To determine the equivalent $V_{\pi}$ of our piezo-optomechanical circuits when considering them to operate as an acousto-optic phase modulator, we use the relationship:
\begin{eqnarray}
V_{\pi} = \pi\frac{V_\text{saw}}{\beta_\text{saw}}
\end{eqnarray}

To determine $V_\text{saw}$, the applied RF voltage that is converted to a propagating surface acoustic wave, we use the resonant dip in the $S_{11}$ spectrum of our IDT to determine the transmitted RF power (in our case -0.25 dB) and assume a 50 $\Omega$ load to determine the corresponding voltage. The modulation index due to the SAW is extracted using the procedure described in the previous section.

\section{Acoustic wave interference from two IDTs}

\begin{figure}[h]
\centerline{\includegraphics[width=0.85\linewidth]{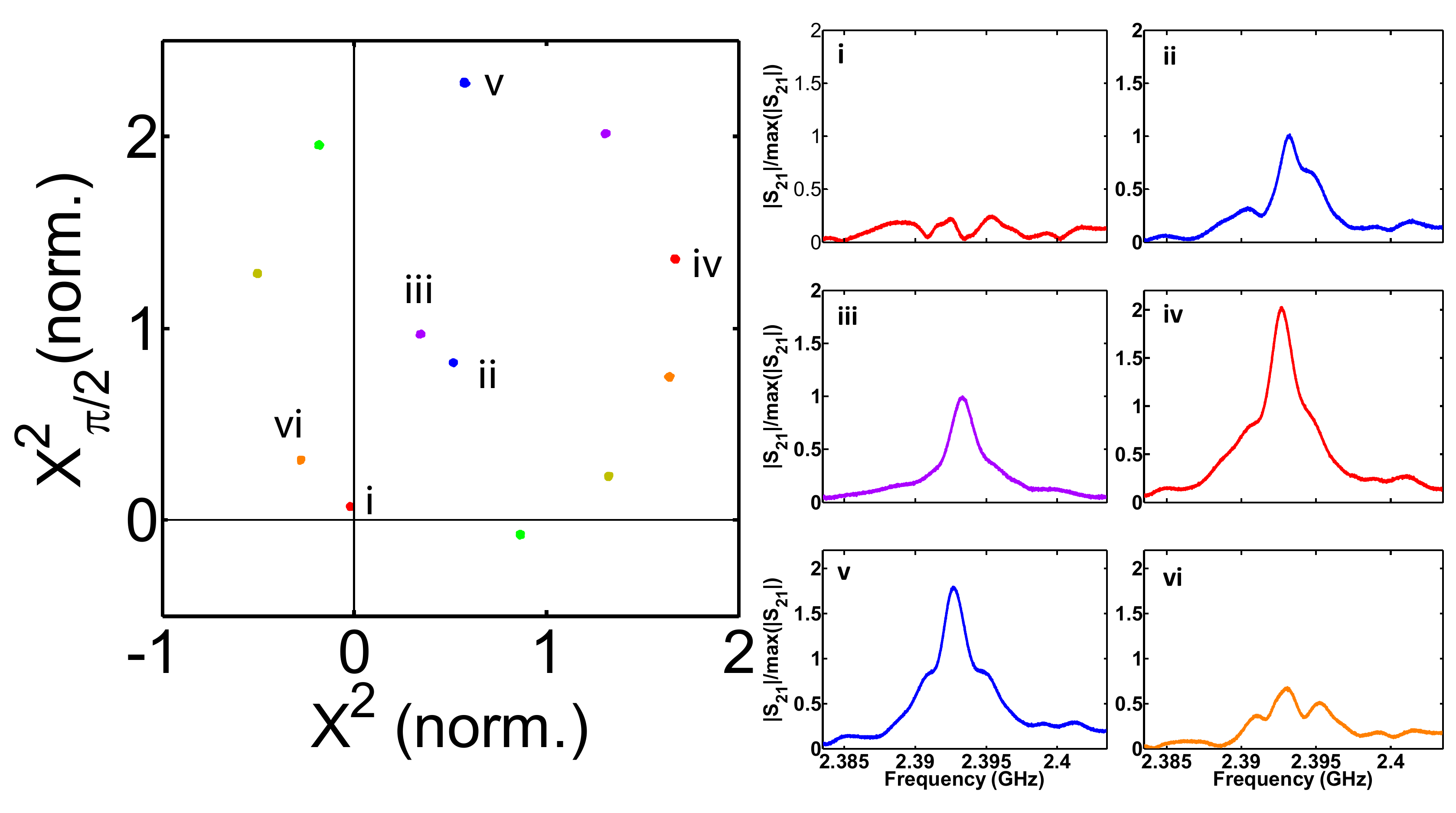}}
 \caption{Phase space diagram for different values of the RF amplitude and phase for a nanobeam optomechanical crystal cavity pumped by two phononic waveguides, each of which is sourced by its own IDT.  The six numbered points, with frequency response curves shown in the side panel, correspond to (i) acoustic wave cancellation, (ii) IDT~1 ON, IDT~2 OFF, (iii) IDT 1 OFF, IDT~2 ON, and (iv, v and vi) Fano lineshape behavior for varying phase difference between IDT~1 and IDT~2.  The measurement bandwidth in these experiments is 200~Hz.}
\label{fig:FigureSI_two_saw_awi}
\end{figure}

The main text discusses acoustic wave interference that occurs when the mechanical mode is driven optically and electrically with equal amplitude and opposite phase. The same phenomenon occurs when the mechanical resonator is pumped by two IDTs through two phononic waveguides with equal amplitudes and opposite phase. To demonstrate this, we use the device shown in Figure 3(a) in the main text, and modify the experimental setup of Figure 2(c) so that the output of the VNA is split in two, with each path fed to its own RF amplifier and then connected to an IDT. The resulting data is presented in Figure~\ref{fig:FigureSI_two_saw_awi}. The two IDTs when turned on alone (ii and iii in the polar plot) result in a regular electro-optic $S_{21}$ measurement as in Figure 2(e), for example. The asymmetry between the spectral responses can be attributed to fabrication-induced asymmetry in the device and the asymmetric coupling induced due to the fiber taper touching down on the device. When the two IDTs are both fed by an RF signal with a varying phase difference, one can sweep out a circular trajectory in phase space. The intersection of this circle with the origin corresponds to the acoustic wave cancellation condition. The mechanical resonator is loaded in a dark state by the destructive interference between the excitation amplitudes of the two IDTs. For an arbitrary phase difference between the two IDTs, the response in general has an asymmetric Fano-like lineshape.

\section{Acoustic coherent population trapping theory}

The equations of motion for the intracavity optical field amplitude and the cavity mechanical displacement amplitude are:

\begin{eqnarray}
\dot{a} = -(i\Delta+{\frac{\kappa_{i}}{2}})a-ig_{0}a(b+b^{\dagger})-\sqrt{\frac{\kappa_{e}}{2}}a_{in}
\label{eq:intracavity_opt_field}
\end{eqnarray}

\begin{eqnarray}
\dot{b} = -(i\Omega_{\text{m}}+{\frac{\gamma_{i}}{2}})b-ig_{0}a^{\dagger}a-\sqrt{\frac{\gamma_{e}}{2}}b_{in}
\label{eq:intracavity_mech_disp}
\end{eqnarray}

\noindent with $a(a^{\dagger})$ representing the annihilation (creation) operator for the intracavity optical field, $b(b^{\dagger})$ the annihilation (creation) operator for the mechanical displacement. $\Delta$ represents the detuning of the control beam from the optical cavity resonance frequency, $\kappa_{i}$ the intrinsic decay rate of the optical cavity, and $\kappa_{e}$ is the extrinsic decay rate (coupling rate) to the waveguide. $\Omega_{m}$ represents the mechanical mode frequency, $\gamma_{i}$ the intrinsic decay rate of the mechanical cavity and $\gamma_{e}$ is the extrinsic decay rate (coupling rate) to the phononic waveguide. $a_{in}$ and $b_{in}$ represent the optical and acoustic field strengths in the photonic and phononic waveguides respectively. $g_{0}$ represents the vacuum optomechanical coupling rate.

We make the following ansatz:

\begin{eqnarray}
a \Rightarrow \alpha = \alpha_{0}+ \alpha_{d+}e^{-i\Omega_{d}t} + \alpha_{d-}e^{i\Omega_{d}t}
\label{eq:intracavity_opt_field_ansatz}
\end{eqnarray}

\begin{eqnarray}
b \Rightarrow \beta = \beta_{0} + \beta_{d+}e^{-i\Omega_{d}t}
\label{eq:intracavity_mech_disp_ansatz}
\end{eqnarray}

{\noindent}Physically, this ansatz correponds to linearizing around the steady state photon ($\alpha_{0}$) and phonon ($\beta_{0}$) amplitudes.  For the mechanical motion:
\begin{eqnarray}
\beta_{d-} = \beta^{*}_{d+}
\end{eqnarray}

{\noindent}Substituting equations~\ref{eq:intracavity_opt_field_ansatz} and~\ref{eq:intracavity_mech_disp_ansatz} into equation~\ref{eq:intracavity_mech_disp} and collecting terms at DC:

\begin{eqnarray}
\beta_{0} = \frac{-ig_{0}|\alpha_0|^2}{i\Omega_{m}+\frac{\gamma_{i}}{2}}
\end{eqnarray}

{\noindent}The true numerator is $-ig_{0}(|\alpha_0|^2+|\alpha_{d+}|^2+|\alpha_{d-}|^2)$ but we neglect the latter two terms under the assumption that $|\alpha_0|^2 >> |\alpha_{d+}|^2+|\alpha_{d-}|^2$.

We can write the incident acoustic displacement as:
\begin{eqnarray}
\beta_{in}  = \beta_{in,0}e^{i\psi}
\end{eqnarray}
{\noindent} wherein $\beta_{in,0}$ is the amplitude of the incident acoustic wave and $\psi$ represents the (variable) phase difference between the RF signal applied to the phase modulator (PM) and the IDT.

{\noindent}Substituting equations~\ref{eq:intracavity_opt_field_ansatz} and~\ref{eq:intracavity_mech_disp_ansatz} into equation~\ref{eq:intracavity_mech_disp} and collecting terms at $e^{-i\Omega_{d}t}$:

\begin{eqnarray}
\beta_{d+}  = \frac{-ig_{0}(\alpha_{0}\alpha^{*}_{d-}+\alpha_{d+}\alpha^{*}_{0})-\sqrt{\frac{\gamma_{e}}{2}}\beta_{in,0}e^{i\psi}}{i(\Omega_{m}-\Omega_{d})+\frac{\gamma_{i}}{2}}
\end{eqnarray}

{\noindent}We can now see that the mechanical resonator's coherent amplitude $\beta_{d+}$ is driven optomechanically, through beating of the pump ($\alpha_{0}$) and probe sidebands ($\alpha_{d+}$ and $\alpha_{d-}$), as well as through the phononic channel by $\beta_{in}$.  When the amplitude of these two terms is equal, the phase $\psi$ can be tuned to achieve cancellation with $\beta_{d+}$=0, corresponding to the acoustic coherent population trapping condition.

{\noindent}Having determined an equation for the mechanical resonator's coherent displacement amplitude, we now consider the system's optical response. The input field can be written as:

\begin{eqnarray}
\alpha_{in}  = \alpha_{in,0}(1+i\frac{\phi}{2}e^{i\Omega_{d}t}+i\frac{\phi}{2}e^{-i\Omega_{d}t})
\end{eqnarray}

{\noindent}where $\phi$ is the modulation index that represents the action of the optical phase modulator on the input optical pump field.

Substituting equations~\ref{eq:intracavity_opt_field_ansatz} and~\ref{eq:intracavity_mech_disp_ansatz} into equation~\ref{eq:intracavity_opt_field} and collecting terms at DC:
\begin{eqnarray}
\alpha_{0}  = -\frac{\sqrt{\frac{\kappa_{e}}{2}}\alpha_{in,0}}{i(\Delta+2g_{0}(\beta_{0}+\beta^{*}_{0}))+\frac{\kappa_{i}}{2}}
\end{eqnarray}

Substituting equations~\ref{eq:intracavity_opt_field_ansatz} and~\ref{eq:intracavity_mech_disp_ansatz} into equation~\ref{eq:intracavity_opt_field} and collecting terms at $e^{-i\Omega_{d}t}$:

\begin{eqnarray}
\alpha_{d+}  = \frac{-2ig_{0}\alpha_{0}\beta_{d+}-\sqrt{\frac{\kappa_{e}}{2}}\alpha_{in,+}}{i(\Delta-\Omega_{d})+\frac{\kappa_{i}}{2}}
\end{eqnarray}

Substituting equations~\ref{eq:intracavity_opt_field_ansatz} and~\ref{eq:intracavity_mech_disp_ansatz} into equation~\ref{eq:intracavity_opt_field} and collecting terms at $e^{i\Omega_{d}t}$:

\begin{eqnarray}
\alpha_{d-}  = \frac{-2ig_{0}\alpha_{0}\beta^{*}_{d+}-\sqrt{\frac{\kappa_{e}}{2}}\alpha_{in,-}}{i(\Delta+\Omega_{d})+\frac{\kappa_{i}}{2}}
\end{eqnarray}

The two sidebands at the output are:

\begin{eqnarray}
|\alpha_{d+,out}| = |\alpha_{d,in+}+\sqrt{\frac{\kappa_{e}}{2}}\alpha_{d+}|
\end{eqnarray}

\begin{eqnarray}
|\alpha_{d-,out}| = |\alpha_{d,in-}+\sqrt{\frac{\kappa_{e}}{2}}\alpha_{d-}|
\end{eqnarray}

In the resolved sideband case (when the pump is blue-detuned), we can derive a condition for observing acoustic wave interference. The relevant equations, in the limit that only the $\alpha_{d-}$ sideband and $\beta_{d+}$ matter, are:

\begin{eqnarray}
\beta_{d+}  = \frac{-ig_{0}(\alpha_{0}\alpha^{*}_{d-})-\sqrt{\frac{\gamma_{e}}{2}}\beta_{in,0}e^{i\psi}}{i(\Omega_{\text{m}}-\Omega_{d})+\frac{\gamma_{i}}{2}}
\label{eq:beta_dplus_eqn}
\end{eqnarray}

\begin{eqnarray}
\alpha_{d-}  = \frac{-2ig_{0}\alpha_{0}\beta^{*}_{d+}-\sqrt{\frac{\kappa_{e}}{2}}\alpha_{in,-}}{i(\Delta+\Omega_{d})+\frac{\kappa_{i}}{2}}
\label{eq:alpha_dminus_eqn}
\end{eqnarray}

Acoustic coherent population trapping corresponds to $\beta_{d+}~=~0$. In this limit, $\alpha_{d-}$ reduces to:

\begin{eqnarray}
\alpha_{d-}  = \frac{-\sqrt{\frac{\kappa_{e}}{2}}\alpha_{in,-}}{i(\Delta+\Omega_{d})+\frac{\kappa_{i}}{2}}
\end{eqnarray}

If we substitute this back in equation~\ref{eq:beta_dplus_eqn}, we get the condition for the amplitude and phase of the propagating acoustic wave input to engineer acoustic coherent population trapping:

\begin{eqnarray}
\sqrt{\frac{\gamma_{e}}{2}}\beta_{in,0}e^{i\psi}= -ig_{0}\alpha_{0}\alpha^{*}_{d-}
\end{eqnarray}

Substituting $\alpha_{d-}$ from equation~\ref{eq:alpha_dminus_eqn} leads us to:
\begin{eqnarray}
\sqrt{\frac{\gamma_{e}}{2}}\beta_{in,0}e^{i\psi}= \frac{-i\sqrt{\frac{\kappa_{e}}{2}}g_{0}\alpha_{0}\alpha^{*}_{in,-}}{-i(\Delta+\Omega_{d})+\frac{\kappa_{i}}{2}}
\end{eqnarray}

\section{Comparison between OMIT, EMIT, and acoustic CPT}

A number of different coherent interference effects have been observed in cavity optomechanical systems, and more recently ~\cite{ref:Agarwal_Huang_OMIT_SI,ref:Weis_Kippenberg_SI,ref:safavi-naeini4_SI}, in piezoelectrically actuated cavity optomechanical systems ~\cite{ref:Bochmann_Cleland_NatPhys_SI,ref:Fong_Tang_uwave_assisted_SI}.  Here, we compare these different effects with those observed in our experiments.  A key conclusion is that while these previous demonstrations have involved interference in the optical domain, the acoustic coherent population trapping effect we have observed is a novel cancellation occuring in the mechanical domain.

\begin{figure}
\centerline{\includegraphics[width=\textwidth,height=\textheight,keepaspectratio]{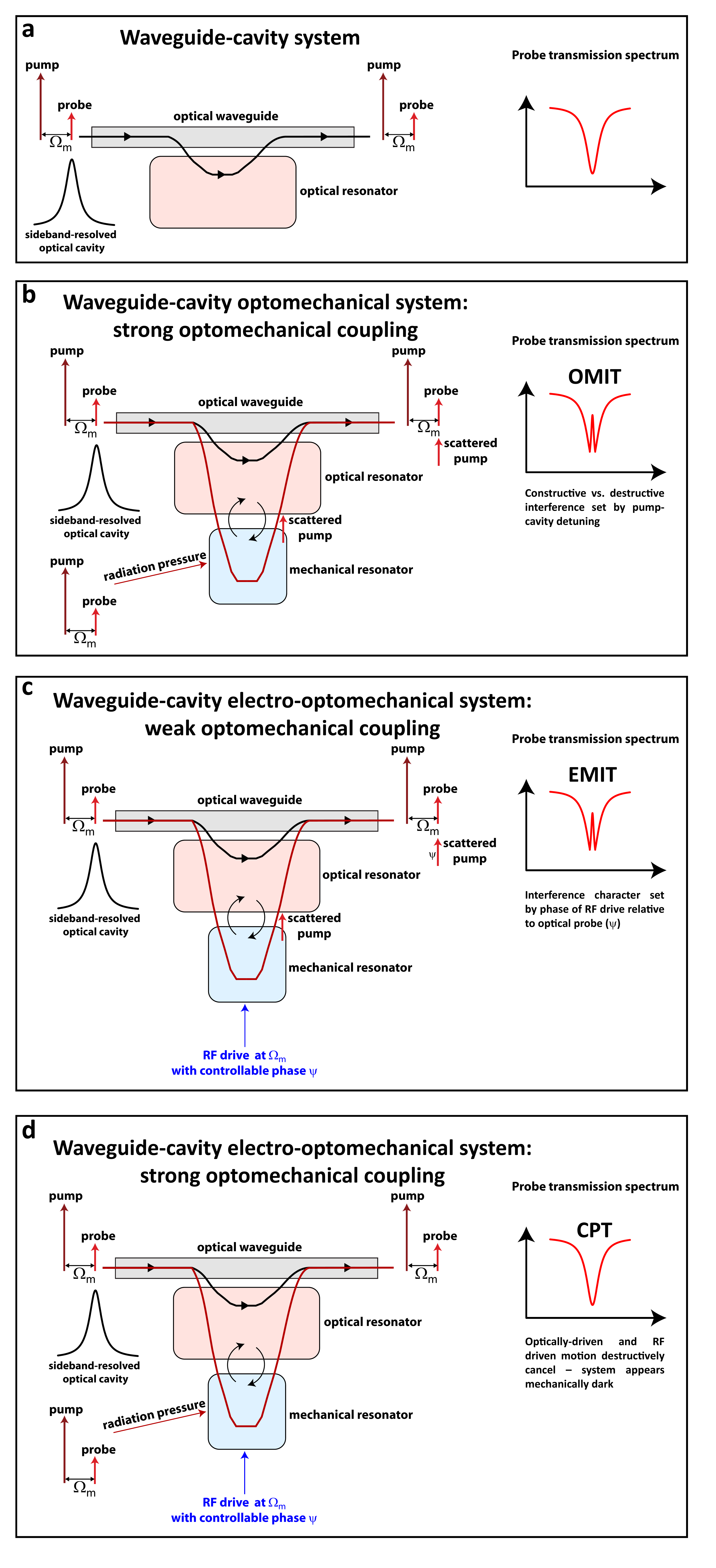}}
 \caption{Illustration of different interference effects observable in the piezo-optomechanical devices. (a) Standard optical waveguide-cavity coupling, in which the mechanical mode does not play a role (optomechanical and electromechanical coupling are set to zero); (b) optomechanically-induced transparency (OMIT); (c) electromechanically-induced optical transparency (EMIT); (d) acoustic coherent population trapping (CPT).  For simplicity, the figures consider the sideband-resolved regime in which only one sideband produced by phase modulation of the optical pump interacts with the system.}
\label{fig:Figure_emit_omit_cpt}
\end{figure}

Figure~\ref{fig:Figure_emit_omit_cpt}  schematically depicts four different physical situations: (a) waveguide-cavity coupling in a system without any optomechanical interaction; (b) optomechanically-induced transparency ~\cite{ref:Agarwal_Huang_OMIT_SI,ref:Weis_Kippenberg_SI,ref:safavi-naeini4_SI}; (c) electromechanically-induced optical transparency (EMIT) or microwave-assisted OMIT ~\cite{ref:Bochmann_Cleland_NatPhys_SI,ref:Fong_Tang_uwave_assisted_SI}; and (d) acoustic coherent population trapping. The pump is detuned one mechanical resonance away from the optical cavity line-center and we monitor the probe transmission by looking at the coherent RF spectrum of the photodetected signal (generated by the beating between the pump and the probe) transmitted past the cavity. We assume that the system is sideband-resolved, so that only one of the sidebands produced by the phase modulator interacts with the cavity.  Our system is sufficiently close to the sideband-resolved regime ($\kappa_{opt}/{2\pi}~\approx~5.2$~GHz, $\Omega_{\text{m}}/{2\pi}~\approx~2.4$~GHz) for this assumption to be largely valid.  In the calculations presented in the main text, the influence of both sidebands was included to model the system as closely as possible.

Starting with a simple side-coupled waveguide cavity system (assuming no optomechanical coupling), an incident optical probe will show a characteristic Lorentzian dip in its transmission spectrum corresponding to the optical cavity resonance (shown in Figure~\ref{fig:Figure_emit_omit_cpt}(a)). If we now turn the optomechanical coupling on, when the frequency difference between the pump and the probe approaches the mechanical resonance frequency, their interference drives the mechanical motion of the cavity and scatters photons from the pump to the probe beam frequency. These additional probe photons can interfere either constructively or destructively (depending on the pump detuning) with the transmitted probe beam (from the regular waveguide-cavity scenario) and leads to the opening of a transparency window in the transmission spectrum (Figure~\ref{fig:Figure_emit_omit_cpt}(b)), a process commonly referred to as optomechanically induced transparency (OMIT)~\cite{ref:Weis_Kippenberg_SI,ref:safavi-naeini4_SI}. It is important to keep in mind that the interference occurs in the optical domain and the mechanical mode is coherently driven (has a non-zero displacement amplitude).

To observe OMIT, the optomechanical coupling rate $g_{0}/2\pi$ has to be sufficiently strong so that the beating between the pump and probe can drive the mechanical motion with sufficient amplitude to scatter photons from the pump to the probe beam frequency. In systems with small $g_{0}/2\pi$, the mechanical motion can be coherently driven by a different mechanism (usually by acoustic waves using the piezoelectric effect) and as long as the RF signal driving the mechanics is derived from the same source as the signal driving the electro-optic phase modulator, the photons scattered by the resonator from the pump will be phase coherent with the probe photons and one can see the induced transparency effect (Figure~\ref{fig:Figure_emit_omit_cpt}(c)). This has been referred to as electromechanically induced transparency or microwave-assisted transparency in the literature~\cite{ref:Bochmann_Cleland_NatPhys_SI,ref:Fong_Tang_uwave_assisted_SI}. In effect, it is not different from OMIT, except the source driving the coherent mechanical motion is electrical rather than optical, and the character of the interference (constructive, destructive, or Fano-like) is set by the relative phase between the electrical drive and optical probe.

Finally, in a system with both strong $g_{0}/2\pi$ and electrical drive, as we have presented in this work, the mechanical resonator can be driven both optically (due to the beating between pump and probe, like in OMIT) or electrically (like in EMIT). By choosing the amplitude and the phase of the electrical drive, one can see either constructive or destructive interference in the mechanical domain (Figure~\ref{fig:Figure_emit_omit_cpt}(d)). Unlike pure OMIT / EMIT, the interference actually occurs in the acoustic domain and in case of destructive interference, the coherent component of the mechanical motion is zero. In this case, the system shows coherent population trapping for phonons analogous to the CPT observed in atomic systems. One can also say that the system has reached a mechanically 'dark' state, and the transmission goes back to that of a bare optical waveguide-cavity system.

\begin{figure}[h]
\centerline{\includegraphics[width=0.85\linewidth]{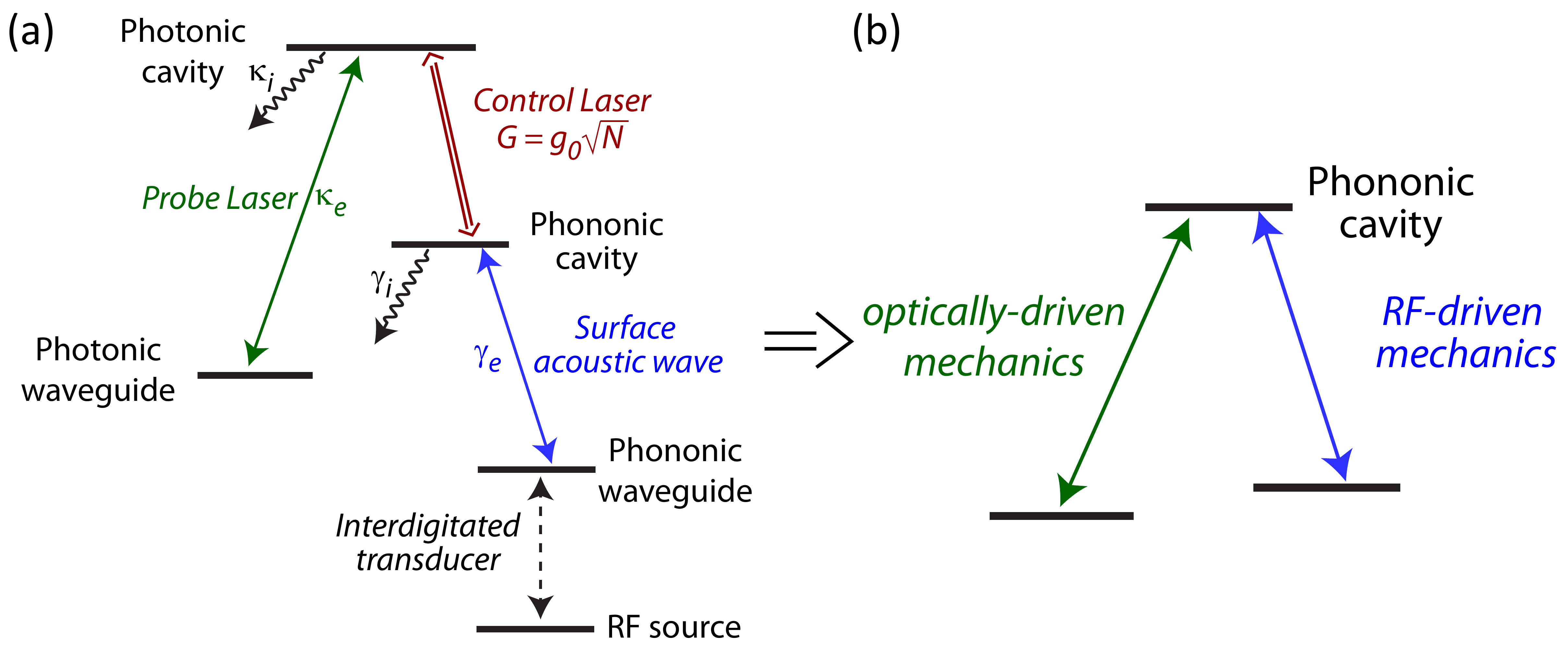}}
 \caption{(a) Schematic level diagram for the piezo-optomechanical circuit, re-displayed from Figure 4(a) of the main text. (b) Effective 3-level $\Lambda$ system for phonons indicating the optomechanically-driven and RF-driven pathways.}
\label{fig:FigureSI_level_diagram}
\end{figure}

The term 'coherent population trapping' was chosen because the overall system, schematically depicted in Figure 4a and re-displayed in Figure~\ref{fig:FigureSI_level_diagram}(a), can be viewed from the perspective of the phononic cavity as an effective three-level system, in which the phononic cavity is populated through either an optomechanically-mediated path or an RF-driven path (Figure~\ref{fig:FigureSI_level_diagram}(b)).  The mechanical dark state - cancellation of the phononic cavity's coherent mechanical motion - occurs when the strength of the two transitions is equal (and phase is opposite).  This is analogous to the situation in atomic physics, where coherent population trapping refers to an interference occuring when the two terms are equal in amplitude, while EIT is used when the control field is much stronger than the probe field ~\cite{ref:Arimondo_CPT_SI,ref:Khan_CPT_EIT_SI}.

Finally, we note that an effect analogous to OMIT, but now consisting of a transparency in the spectrum of propagating acoustic probe phonons due to coupling to intracavity photons, should also be observable in this system.

\section{Absence of perfect cancellation in coherent population trapping}

The inability to achieve perfect cancellation (a flat transmission spectrum in either the cyan curve from Figure 4(c), or Figure 4(e), curve ii) can be attributed to the difference between the optomechanical ($K_\text{om}$) and electromechanical ($K_\text{em}$) transduction spectra for different mechanical modes in the optomechanical cavity. To illustrate this, we consider the difference between the OMIT response wherein we are primarily measuring the optomechanical transduction coefficient and the electro-optic $S_{21}$ measurement wherein we measure the product of the two:

\begin{eqnarray}
PSD_{OMIT}\propto K_\text{om}(\Omega)
\end{eqnarray}

\begin{eqnarray}
PSD_{EO-S21}\propto K_\text{em}(\Omega)K_\text{om}(\Omega)
\end{eqnarray}

The optomechanical transduction ($K_\text{om}$) is calculated by an overlap integral between the localized optical and mechanical modes whereas the electromechanical transduction is estimated by an overlap integral between the localized mechanical mode and the surface acoustic wave incident on the cavity. When a surface acoustic wave is incident on the cavity, it excites a superposition of all the mechanical modes of the cavity which lie within the given frequency range and the corresponding mode amplitudes are given by the corresponding transduction coefficients.

\begin{figure}[h]
\centerline{\includegraphics[scale=0.5]{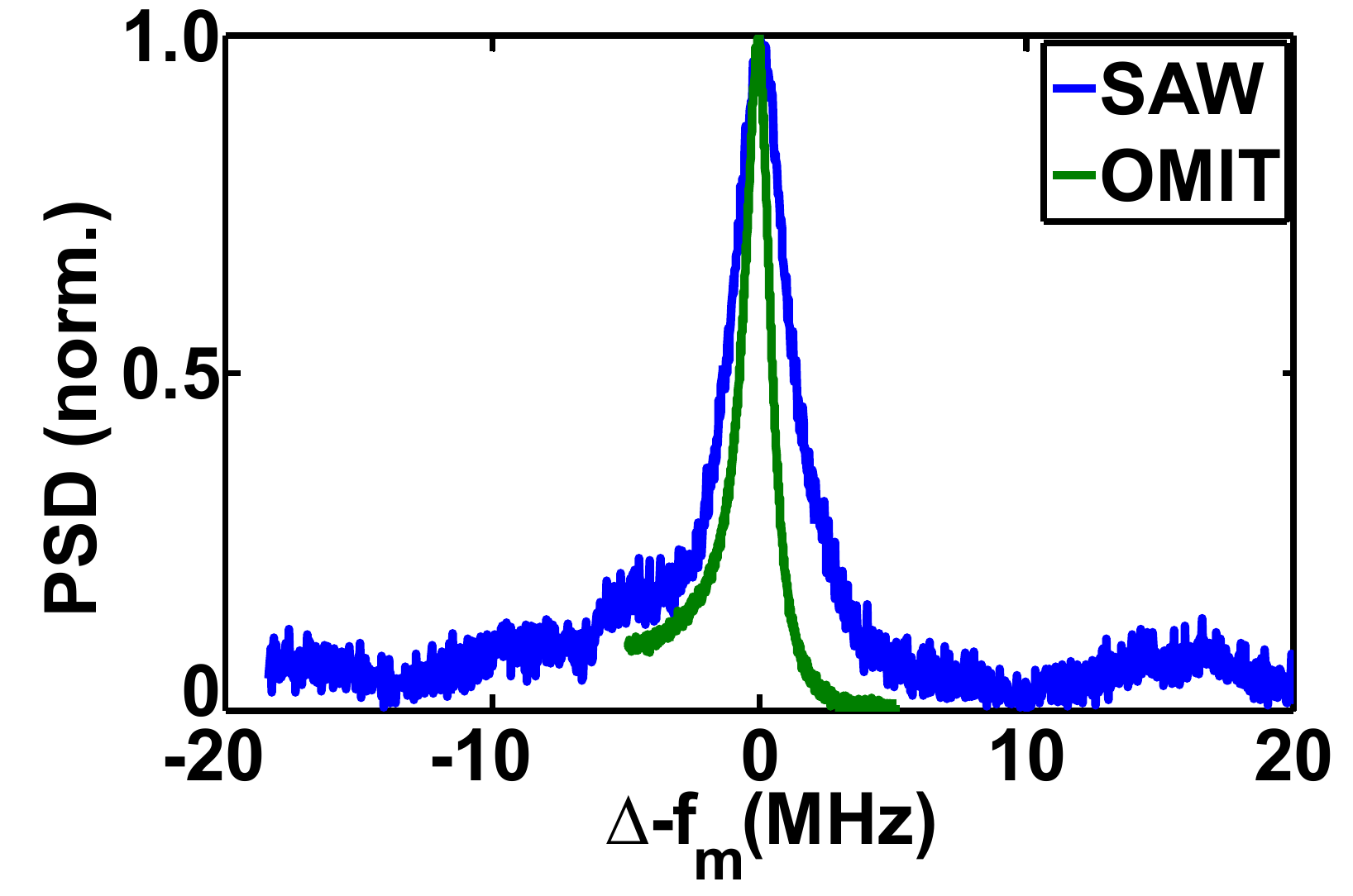}}
 \caption{(a) Normalized probe transmission for a nanobeam optomechanical cavity when the mechanical mode is coherently driven (i) optically (OMIT, green curve) and (ii) acoustically, through a surface acoustic wave excited by the IDT (SAW, blue).}
\label{fig:FigureSI_saw_omit_diff}
\end{figure}

This helps us understand why the two spectra appear different (shown in Figure~\ref{fig:FigureSI_saw_omit_diff} (a)). In an OMIT measurement, the only mode that is excited with significant amplitude is a mode with strong optomechanical coupling (in this case, the localized breathing mode) and hence, the power spectral density has a Lorentzian lineshape corresponding to a single mode. When a propagating acoustic wave is incident on the cavity, other modes which might have small optomechanical coupling but significant electromechanical coupling (due to a strong overlap with the propagating acoustic wave) also appear in the measured transmission spectrum. Since the acoustic wave interference condition requires both the modes to have similar optomechanical and electromechanical transduction for the cancellation to occur, only the breathing mode is cancelled while the other excitations are not. This leads to the appearance of residual peaks and a non-flat background in the transmission spectrum. (Note: we believe these additional mechanical modes occur due to fabrication imperfections in the dimensions of the elliptical holes in the nanobeam cavity).

\section{Acoustic wave interference for a nanobeam cavity with two closely spaced mechanical modes}

As illustrated in the previous section, different closely spaced mechanical modes can have vastly different optomechanical and electromechanical coupling coefficients. This enables us to demonstrate conclusively that the acoustic wave interference condition shown in Figure 4(c) in the main text occurs in the mechanical domain due to a cancellation between the optomechanical and electromechanical drive terms.

\begin{figure}[h!]
\centerline{\includegraphics[width=\linewidth]{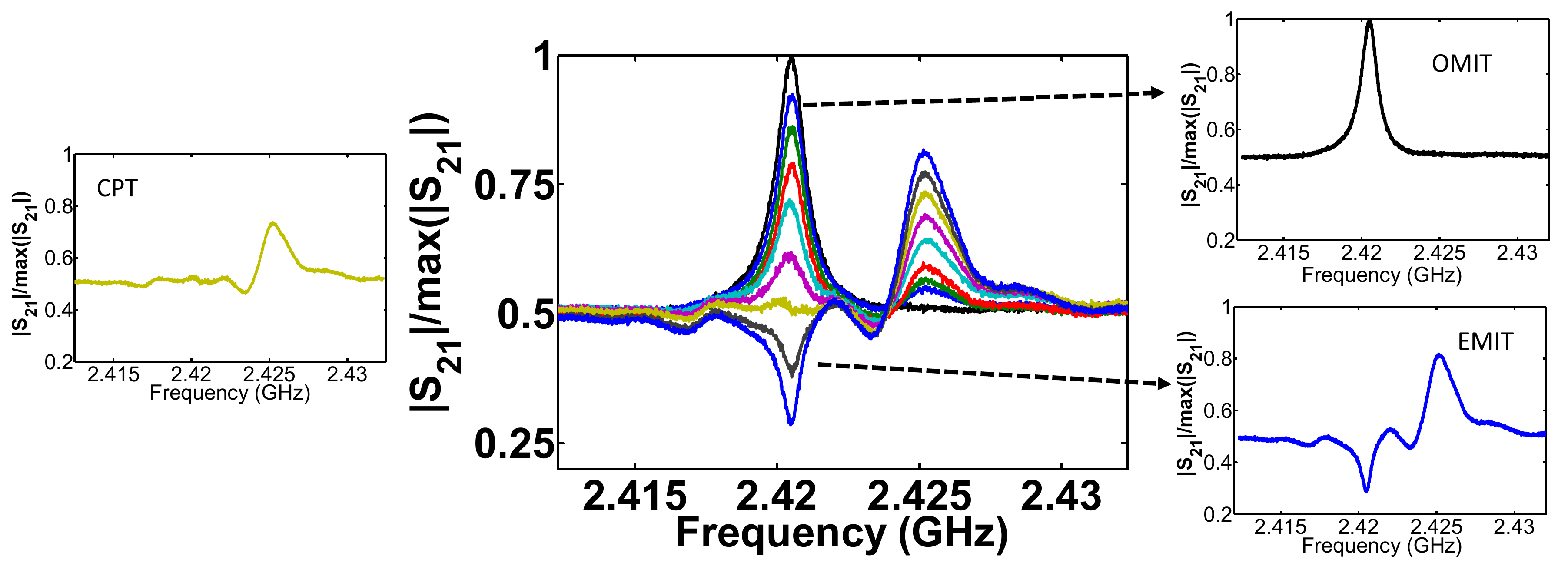}}
 \caption{Normalized probe sideband transmission amplitude as a function of increasing RF power to the IDT. The lower frequency mode (with strong optomechanical coupling) shows a clear signature of acoustic coherent population trapping whereas the higher frequency mode does not. The OMIT and EMIT cases, corresponding to zero and maximum RF power applied to the IDT, respectively, are shown separately for clarity on the right, while the acoustic coherent population trapping case (CPT) is shown separately on the left.}
\label{fig:FigureSI_two_mode_awi}
\end{figure}

Figure~\ref{fig:FigureSI_two_mode_awi} shows an optomechanical nanobeam cavity with two closely spaced mechanical modes, one the breathing mode which has both strong optomechanical and electromechanical coupling and a second mode which has strong electromechanical coupling but weak optomechanical coupling. We use the exact same procedure as in Figure 4(c) to probe the device. Starting with no RF power applied to the IDT, we see a single Lorentzian peak corresponding to OMIT (black curve) wherein only the mode with strong optomechanical coupling is visible. As we increase the RF power to the IDT keeping the phase set for acoustic wave interference, we see that the first mode decreases in amplitude whereas the second (higher frequency) mode rises. This opposite behaviour can be explained by noting that the first mechanical mode (breathing mode) is being driven by two terms, the optical drive term due to the beating between the carrier and the phase modulated sideband and the acoustic term due to the surface wave incident on the cavity. In contrast, the second mode is primarily driven by the propagating acoustic mode. Thus, the first mode clearly demonstrates the acoustic cancellation effect when the two driving terms are comparable in amplitude but opposite in phase whereas the second mode does not show any acoustic wave interference effect because there is only one driving term.

\end{document}